\documentclass[fleqn,usenatbib]{mnras}

\usepackage{newtxtext,newtxmath}

\usepackage[T1]{fontenc}

\DeclareRobustCommand{\VAN}[3]{#2}
\let\VANthebibliography\thebibliography
\def\thebibliography{\DeclareRobustCommand{\VAN}[3]{##3}\VANthebibliography}

\usepackage{graphicx}	%
\usepackage{amsmath}	%
\usepackage{siunitx}

\graphicspath{{./}{figures/}}

\title[Ensemble Mapping the Inner Structure of Luminous Quasars]{Ensemble Mapping the Inner Structure of Luminous Quasars}

\author[L. Wu et al.]{
Liang Wu$^{1,2}$\thanks{E-mail: wul@mail.ustc.edu.cn}
Jun-Xian Wang$^{1,2}$\thanks{E-mail: jxw@ustc.edu.cn}
Hao-Chen Wang$^{1,2}$
Wen-Yong Kang$^{1,2}$
Wei-Da Hu$^{1,2}$
\newauthor{
Ting-Gui Wang$^{1,2}$
and Hui-Yuan Wang$^{1,2}$
}
\\
$^{1}$CAS Key Laboratory for Research in Galaxies and Cosmology, Department of Astronomy, University of Science and Technology of
China, Hefei, Anhui 230026, \\China\\
$^{2}$School of Astronomy and Space Science, University of Science and Technology of China, Hefei 230026, China
}

\date{Accepted XXX. Received YYY; in original form ZZZ}

\pubyear{2022}

\begin{document}
\label{firstpage}
\pagerange{\pageref{firstpage}--\pageref{lastpage}}
\maketitle

\begin{abstract}
A simple prediction of the well-known unification model of active galactic nuclei is that a  sample of sources should exhibit an anti-correlation between the solid angle of the dusty torus and of the ionization cone (as the sum of them shall equal 4$\pi$), which however has never been detected. In this work, we analyze the correlation between [OIII] 5007 narrow emission line equivalent width and $L_{\rm IR}(\lambda)/L_{\rm bol}$ for a large sample of luminous quasars. For the first time, we detect a clear intrinsic anti-correlation between them, which immediately verifies the torus/ionization-cone geometry in luminous quasars. More interestingly, the anti-correlation significantly weakens with increasing wavelength from $\sim$ 2 to 12 \micron, and disappears at $\sim$ 12 \micron. Simulations show a cool dust component (in addition to equatorial torus) with its strength positively correlating with the solid angle of the ionization cone is required to explain the observations. This shows that the polar dust seen in nearby active galaxies also exists in luminous quasars, with its contribution to total dust emission increasing with $\lambda$ (from $\sim$ 2 to 12 \micron) and reaching between 39\%--62\% ({model dependent}) at rest frame 12 \micron. Our findings provide a unique approach to map the otherwise spatially unresolvable inner structure of quasars.
\end{abstract}


\section{Introduction}

The key gradient in the renowned unification scheme of active galactic nuclei is the dusty toroidal structure in the equatorial plane, namely the torus  \citep[e.g.][]{antonucci_unified_1993,Urry1995}. 
The torus obscures radiation from the central engine (re-emits in infrared), and results in the bi-polar ionization cones where narrow emission lines are produced (i,e, the narrow line region, NLR) as detected in many local AGNs and quasars \citep{Evans1991, muller_nlr_2011, storchi_nlr_2018}. 
The torus is widely believed to be clumpy \citep[e.g.][]{krolik_clumpy_1988,nenkova_agn_2008}, with
the innermost part \citep[named hot dust,][]{barvainis_hot_dust_1987} emitting at sublimation temperature T $\sim 500 - 2000$K, and outer part with lower temperature.  
However, while the equatorial dusty structure has been resolved in some nearby active galaxies \citep[e.g.][]{Jaffe1996, leftley_resolving_2021}, they are far from resolvable in most quasars which are far more distant. 

Apart from the toroidal torus, recent high-resolution mid-infrared observations have revealed polar elongated dust emission ubiquitously in nearby AGNs, on scales up to tens or larger parsecs \citep[e.g.][]{Asmus2016,Lopez2016, asmus_new_2019}\footnote{ 
Alternatively, \cite{Nikutta2021} proposed that, under specific conditions, clumpy equatorial tori could exhibit morphologies with significant polar elongation in the mid-infrared.}. Such polar dust, tentatively associated with dusty winds \citep{Honig2017, Stalevski2019}, 
or the dusty NLR \citep[e.g.][]{Netzer1993,Schweitzer2008, Mor2009,Mor2012,Asmus2016},
is expected to be cooler than the hot dust at smaller scales, and could dominate the mid-infrared emission in AGNs.

Within the unification scheme, one can easily foresee a consequence, which however has never been detected, that a sample of sources should exhibit an anti-correlation between the solid angle of the torus and of the ionization cone, i.e., sources with larger solid angle of the torus shall have smaller solid angle of the ionization cone (as the sum of them shall equal 4$\pi$). 
The ratio of monochromatic or integrated infrared luminosity to bolometric luminosity has widely been adopted to quantify the covering factor of the dusty torus in AGNs
\cite[e.g.][]{Maiolino2007,
roseboom_ir-derived_2013,
ma_covering_2013,
wang_outflow_2013,
netzer_revisiting_2015,stalevski_dust_2016}. 
This approach is further supported by the fact that for clumpy torus, the IR SEDs barely change with viewing angle, as long as for type 1 AGNs (viewed along dust-free line of sight, \citealt{stalevski_torus_2012})\footnote{
However note recent works suggested that the connection between $L_{\rm IR}/L_{\rm bol}$ and torus covering factor is likely non-linear \citep{stalevski_dust_2016, toba_how_2021} because of the possible anisotropic radiation from the accretion disk. Such effect would be discussed in \S\ref{sec:anisotropy}.}.

While there are many factors which could affect the observed $L_{\rm IR}/L_{\rm bol}$ in AGNs (see discussion in \S\ref{sec:discussion}),
clearly, for individual sources in a large sample of AGNs, the ratio of torus IR emission to bolometric luminosity is expected to positively correlate with the solid angle of the torus.  
Contrarily, the opposite would be true for the polar dust, whose radiation is expected to correlate with the solid angle of the ionization cone.
Meanwhile, [OIII] 5007 $\si{\AA}$ is the most prominent narrow emission line in the optical spectra of quasars and AGNs, with its luminosity expected to be proportional to the covering factor of the narrow line region, therefore, of the ionization cone. 

In this work for simplicity we define CF($\lambda$) as the ratio of monochromatic infrared luminosity at wavelength $\lambda$ to bolometric luminosity.
We aim to explore the correlation between $L_\mathrm{IR}$($\lambda$)/$L_{\mathrm{bol}}$ and  O[III] EW, two quantities served as proxies of the solid angles of the dusty structure and the ionization cone respectively, of a large sample of luminous quasars.  We expect to detect negative correlation between two quantities if the observed $L_\mathrm{IR}$($\lambda$) is torus dominated, and contrarily negative correlation if polar dust dominated.
By exploring such correlation and its dependence on $\lambda$, %
we could uniquely and ensemble map the otherwise spatially unresolvable inner structure of quasars.
The paper is organized as follows: In \S2, we illustrate the sample selection and the measurements of $L_\mathrm{IR}$($\lambda$)/$L_{\mathrm{bol}}$. In \S3, we describe the correlation analyses and show the results. We discuss the results in \S4 and summarize in \S5.

\section{The Sample and the measurements of $L_\mathrm{IR}$($\lambda$)/$L_{\mathrm{bol}}$}

\subsection{The Sample}\label{sec:S2.1}

We start from the Sloan Digital Sky Survey \citep{sdss} data release 7 quasar catalogue \citep{schneider_sloan_2010}. 
The multiband photometry, including SDSS \citep{fukugita_sloan_1996}, 2MASS\footnote{UKIDSS \citep{ukidss} is the successor to 2MASS, with one more band (Y band), and better data quality. However, it currently only covers 4000 deg$^2$ of the sky and only 217 sources (11.5\%) in our sample can be found in the catalog of UKIDSS. 
} \citep{carpenter_color_2001} and WISE \citep{wright_wide-field_2010}, and SDSS spectral fitting results (line EW, 
and supermassive black hole mass) were taken from \cite{shen_catalog_2011}.
We exclude quasars with redshift $z$ $>$ 0.797 to ensure the coverage of [OIII] 5007 emission line in SDSS spectra. %
To avoid contamination from the host galaxy, following  \cite[][]{roseboom_ir-derived_2013} we focus on the 1953 luminous quasars with bolometric luminosity from  \cite{shen_catalog_2011}\footnote{The bolometric luminosity provided by \cite{shen_catalog_2011} was converted from $L_{5100\AA}$ (measured from SDSS spectral fitting) with a constant bolometric correction factor of 9.26.} above $10^{46}$ erg/s.
We select 1891 sources with 2MASS and WISE detections and the signal to noise ratio (S/N) in all WISE bands greater than 3. Note that since we focus on luminous quasars, utilizing the latest data release (DR16) of SDSS quasar catalog \citep{sdss_dr16}
could only increase the sample size by $\sim$ 20\%. 
238 sources (12.6\%) in our sample have radio loudness $R=F_{\nu,6\si{cm}}/F_{\nu,2500\si{\angstrom}}$ greater than 10, which can be regarded as radio loud \citep{Kellermann1989}.
Excluding them from this study would not significantly alter the results presented in this work.

\subsection{Measuring $L_\mathrm{IR}$($\lambda$)/$L_{\mathrm{bol}}$}\label{sec:S2.2}

In literature, the dust emission $L_\mathrm{IR}$($\lambda$) is commonly derived through SED-fitting the broadband (UV/optical to infrared) photometric data, adopting various templates of disc and dust emission, such as the popular hot+cold dust model \citep[e.g.][]{roseboom_ir-derived_2013}, and X-Cigale \citep{Yang2020}. Clearly, as the broadband SED is sparsely sampled in wavelength, these physical models would yield significantly model/template dependent measurements, and properly decomposing the various dust components thus would be highly challenging. In this work we aim to perform ensemble analyses without the need to decomposing the dust components in individual sources (thus without prior assumptions of the dust templates, which is one unique feature of this work). To do so, we simply interpolate the SED to measure $L_\mathrm{IR}$($\lambda$) in a template-indepdent manner.

\begin{figure}
    \centering
    \includegraphics[width=0.5\textwidth]{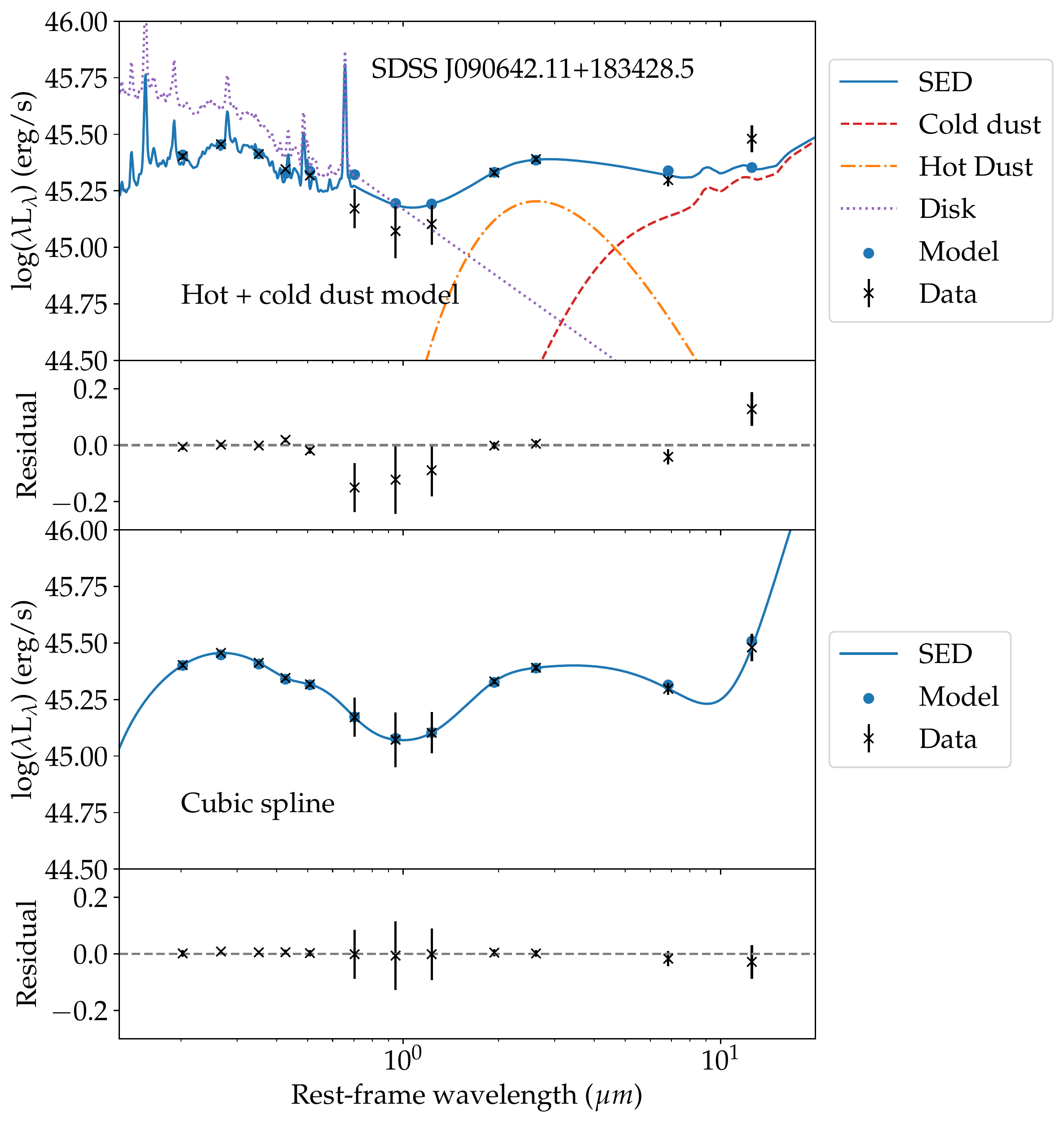}
    \caption{
        Example SED fitting of a randomly chosen source SDSS J090642.11+183428.5, using the hot+cold dust model (upper), and the cubic spline fit (lower).
        Note as the model luminosity in a given band is calculated by convolving the model spectrum over the response curve of each band, the model SED data points do not necessarily fall right on the model spectra.
    }
    \label{fig:example}
\end{figure}

\begin{figure}
    \centering
    \includegraphics[width=0.5\textwidth]{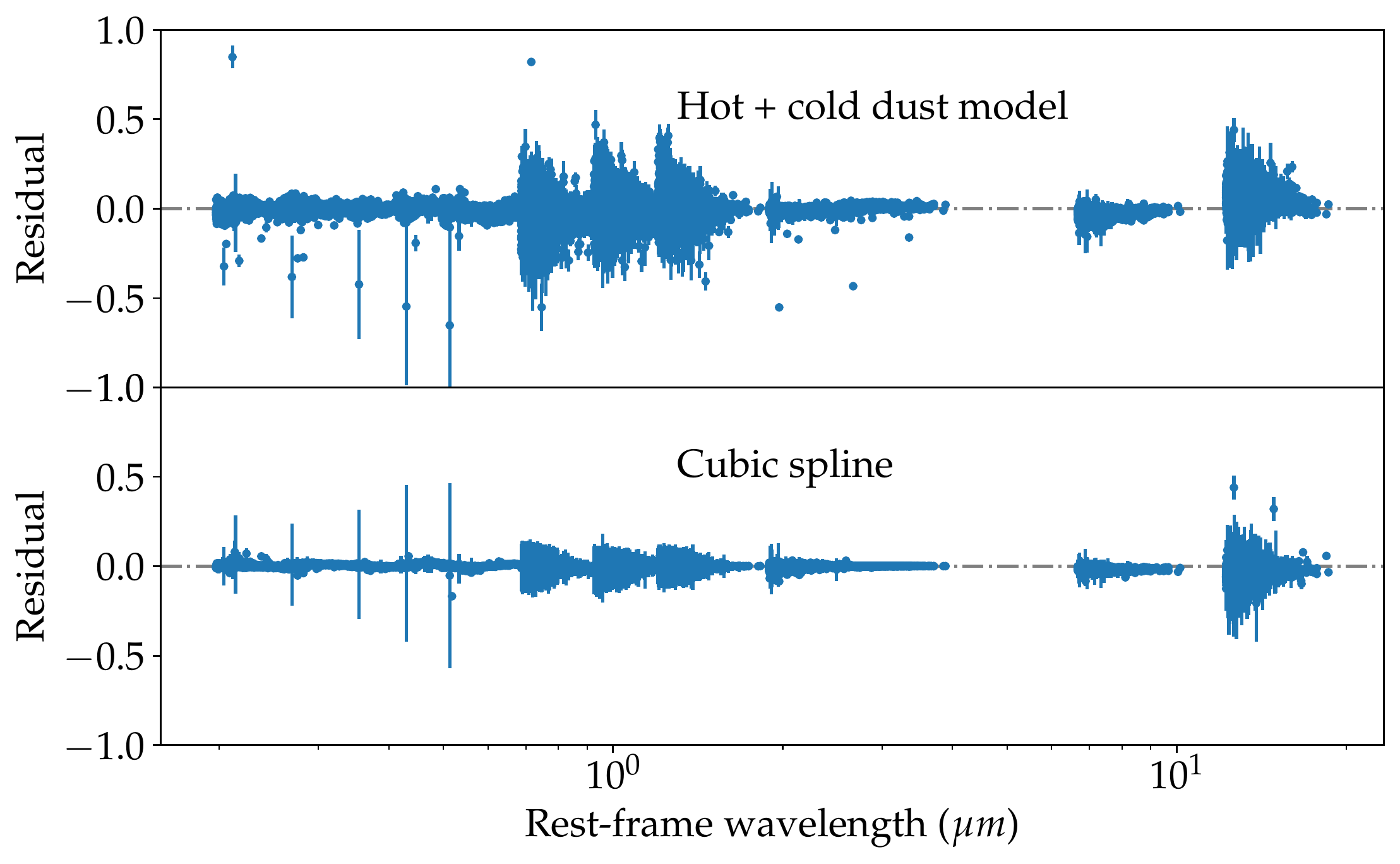}
    \caption{
        Residuals from all bands of all sources of hot+cold dust model (upper), and cubic spline fit (lower). 
    }
    \label{fig:residual}
\end{figure}

Before that, we first show the SED-fitting with the hot+cold dust model of \cite{roseboom_ir-derived_2013}, the measurements of accretion disc component of which could be used in this work.
In the model, the template of the accretion disk component consists of two parts connected at $0.7~\micron$:
the $0.05 - 0.7~\micron$ radio quiet quasar template from \cite{shang_new_2011}
and a power-law of $\lambda L_{\lambda} \propto \lambda^{-1}$ at $\lambda~>~0.7~\micron$. 
Hot dust refers to the dust at or close the sublimation radius which mainly radiate in near-IR
and is modeled with a single black body (with temperature ranging from 500 K to 2000 K).
The 3rd component is to describe the colder dust emission (presumably in the dusty torus, but see later for discussion).
Three cold dust templates from \cite{nenkova_agn_2008} are adopted \citep[see table 2 in][for model parameters]{roseboom_ir-derived_2013}, and the one which yields the least $\chi^2$ among the three is selected for an individual quasar.
Extinction with a SMC-like extinction curve \citep{gordon_quantitative_2003} is considered. Negative extinction is allowed during the fitting to handle a significant fraction of optical/UV SEDs which are bluer than the disc template (as the disc template itself could have been slightly reddened).
Contributions of the host galaxies are not considered.
We plot in Fig. \ref{fig:example} an example fit to an individual quasar.
For most sources, the model we adopted yields reasonably good fits.
The yielded $\chi^2$ (reduced $\chi^2$)
has a median value of 28.9 and a mean of 43.1 for a degree of freedom of 7.\footnote{The median and mean $\chi^2$ we derived are about 2 times larger than those reported by \cite{roseboom_ir-derived_2013}, the reason of which is yet unclear. In fact, if fitting the same SED data points of \cite{roseboom_ir-derived_2013}, i.e., SDSS + UKIDSS + WISE, instead of SDSS + 2MASS + WISE adopted in this work, we obtain even larger $\chi^2$. It is likely that \cite{roseboom_ir-derived_2013} has added extra photometric errors to SED data points.} 
However, there are still a small amount of sources that are poorly
fitted. One possible reason is strong extinction, which may not be well fitted with a simple SMC extinction curve. Strong extinction 
can also affect the measurements of other SED parameters.
Thus we drop sources with $\chi^2>50$ or A$_V$ $>$ 0.5 from following studies in this work, although keeping them would not alter the main results presented in this work. 
Eventually, 1459 sources are left.
For each source we derive $L_{\mathrm{bol}}$ from the monochromatic luminosity of disk template at 5100 $\si{\angstrom}$ adopting a constant bolometric correction factor of 9.26 (the same as that utilized in \citealt{shen_catalog_2011}).
We also calculate the total dust emission through integrating the best-fit model within 1 -- 10 \micron, and derive the corresponding $L_{IR}^{dust}(1-10\micron)$/$L_{\rm bol}$ (hereafter CF0).
We plot in Fig. \ref{fig:residual} the SED fitting residuals for all sources. While the hot+cold dust model yields reasonable fit to UV/optical, clear systematical residuals are seen in infrared.

We then perform cubic spline interpolation to fit the SED (see Fig. \ref{fig:example} and \ref{fig:residual}), which well desrcibe the data and yield no clear systematical residuals. The best-fit SED shape derived through cubic spline interpolation however appears signicantly different from that of the hot+cold dust model at longest wavelengths (see Fig. \ref{fig:example}). To furture examine whether the results in this work is sensitive to the selected interpolation function, we also perform linear interpolation (the most simple form of interpolation).  

Hereafter we proceed with $L_\mathrm{IR}$($\lambda$) measured using cubic spline and linear interpolation, which are simple and template independent. 
Since the hot+cold dust model yields reasonable fit to UV/optical data (see the upper panel in Fig. \ref{fig:residual}), we directly take its output $L_{\mathrm{bol}}$ (converted from $L_{5100\si{\AA}}$ of the best-fit disc template) to calculate $L_\mathrm{IR}$($\lambda$)/$L_{\mathrm{bol}}$.
We further note that, 
if we assume a AGN disc component with a given template (as adopted in the hot+cold dust model), %
$L_\mathrm{IR}$($\lambda$)  = $k_\lambda L_\mathrm{bol} + L_\lambda^\mathrm{dust}$, 
where $k$ at given infrared wavelength is simply a constant for different sources. 
Therefore we do not need to subtract the disc component from the interpolated $L_\mathrm{IR}$($\lambda$), as it does not affect the correlation analyses presented below.

\section{Correlation analyses}\label{sec:S3}

\begin{figure}
    \centering
    \includegraphics[width=0.45\textwidth]{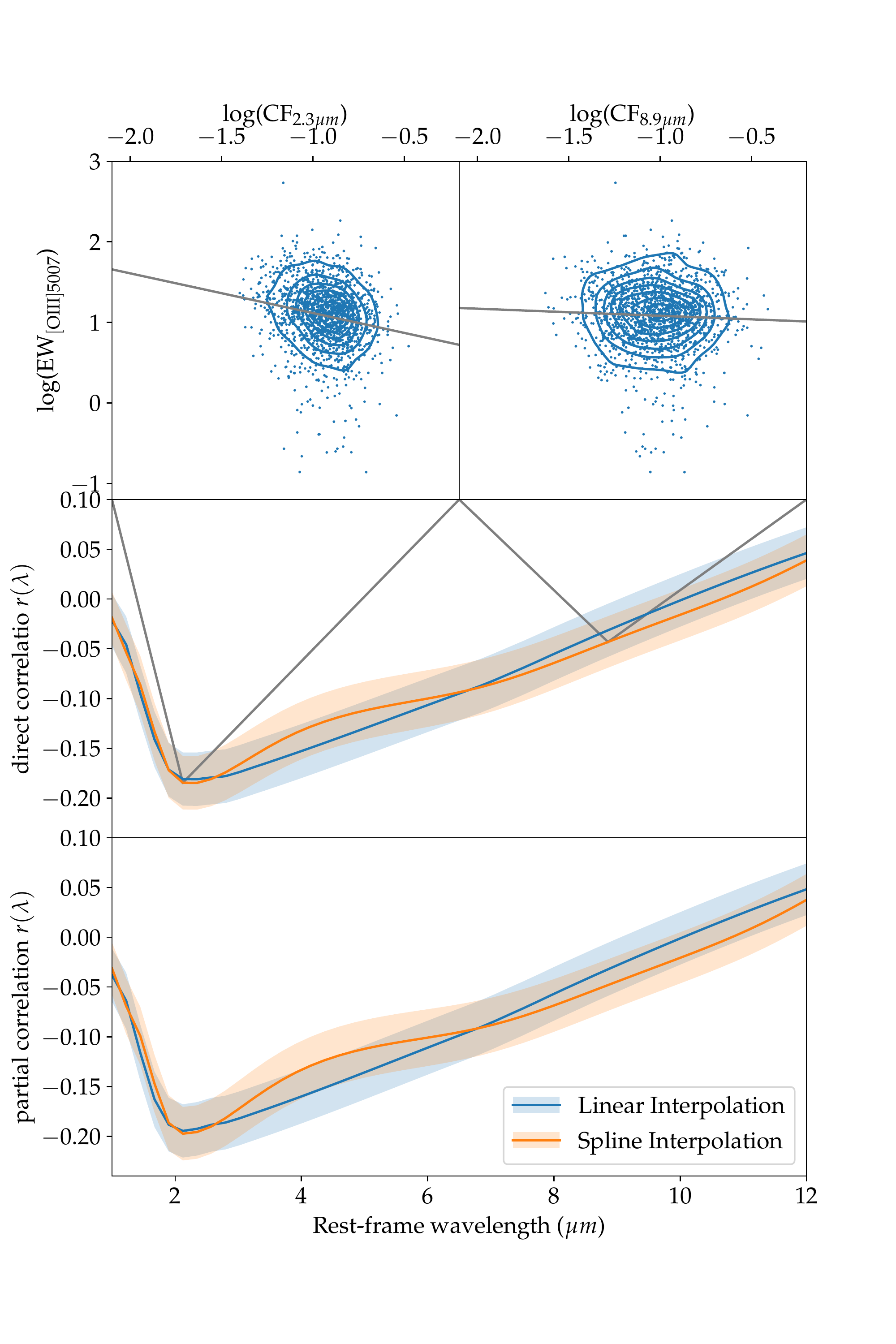}
    \caption{
        Direct (middle panel) and partial correlation (lower panel) coefficient $r$ between $\mathrm{EW}_{\mathrm{[OIII]5007}}$ and CF($\lambda$) as a function of $\lambda$  over rest-frame 1--12 $\micron$. Shadows plot the
        1$\sigma$ confidence errors derived through bootstrapping the sample. 
        To directly demonstrate the correlations, in the upper panels we plot $\mathrm{EW}_{\mathrm{[OIII]5007}}$ versus CF($\lambda$) (calculated with spline interpolation for example)
        and the best-fit linear regression at 2.3 and 8.9 $\micron$, respectively.
    }
    \label{fig:curve}
\end{figure}

\begin{figure}
    \centering
    \includegraphics[width=0.5\textwidth]{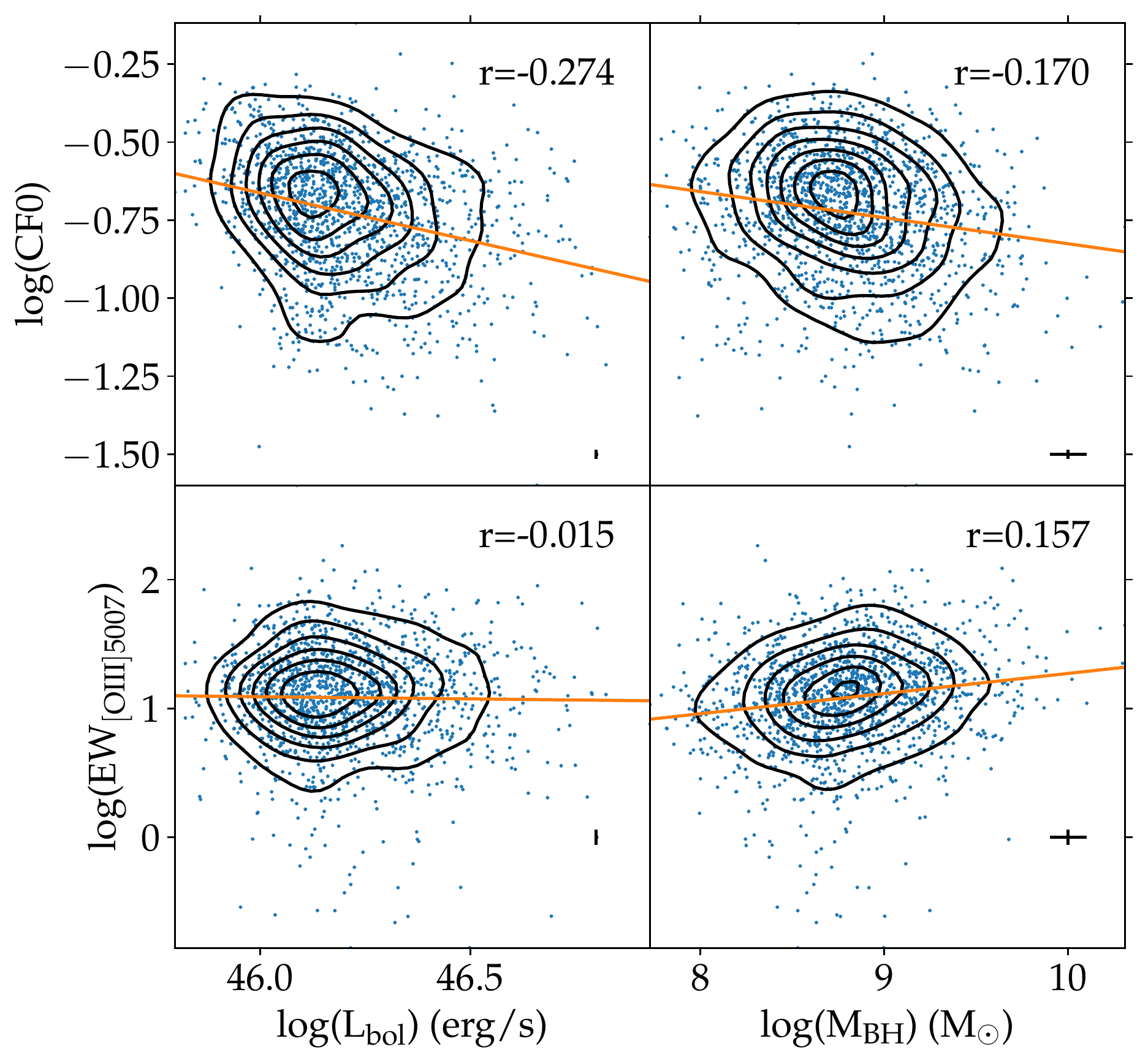}
    \caption{
        $\mathrm{EW}_{\mathrm{[OIII]5007}}$  and CF0 versus bolometric luminosity $L_{\mathrm{bol}}$ and
        black hole mass $M_{\mathrm{BH}}$, respectively. 
        Simple linear regressions are performed and shown as solid lines. 
        The error bars in the lower right corner of each panel mark the median 1$\sigma$ errors of the corresponding parameters for the sample.
        The uncertainties in SMBH mass and $\mathrm{EW}_{\mathrm{[OIII]5007}}$ are from \protect\cite{shen_catalog_2011},
        and those for CF0 and $L_{\rm bol}$ are from SED fitting obtained through adding random Gaussian photometric errors to SED data points. 
        Clearly the statistical errors of the parameters are small comparing with the scatter in the plot, and are ignored in correlation analyses (hereafter the same). 
        The correlation coefficients $r$ between the quantities are labelled. 
    }
    \label{fig:cors}
\end{figure}

We perform Pearson correlation between $\mathrm{EW}_{\mathrm{[OIII]5007}}$ and 
CF($\lambda$)
of our sample and present the resulted correlation coefficient $r$($\lambda$) in Fig. \ref{fig:curve}. 
We estimate the uncertainties in $r$($\lambda$), which are 
also useful to demonstrate the statistical significance of the correlation, through bootstrapping the quasar sample. %
The upper wavelength bound of $r$($\lambda$) is set at 12 $\micron$, corresponding to the rest frame wavelength that WISE W4 (22.19$\micron$) could probe for the highest redshift quasars (z=0.797) in the sample.  

Before we interpret the derived $r$($\lambda$), we note that
both $\mathrm{EW}_{\mathrm{[OIII]5007}}$ and CF($\lambda$) may correlate with fundamental parameters of quasars, including bolometric luminosity $L_{\mathrm{bol}}$, supermassive black hole mass $M_{\mathrm{BH}}$ (and also the Eddington ratio), and such correlations may produce artificial direct correlation between $\mathrm{EW}_{\mathrm{[OIII]5007}}$ and CF($\lambda$).

Note for each source we have two $L_{5100}$ (and $L_{\rm bol}$ = 9.26 * $L_{5100}$), one obtained through SDSS spectral fitting \citep{shen_catalog_2011}, and another from SED fitting to photometric data (\S\ref{sec:S2.2}).
Two $L_{\rm 5100\AA}$ could be different due to the intrinsic variability since the SDSS photometric and spectroscopic observations were not taken simultaneously. 
Hereafter we adopt the SED-fitting based $L_{\rm 5100\AA}$ and $L_{\rm bol}$, though simply utilizing $L_{\rm bol}$ from \cite{shen_catalog_2011} would not alter the key results presented in this work. 

In Fig. \ref{fig:cors} we draw scatter plots presenting the dependence of $\mathrm{EW}_{\mathrm{[OIII]5007}}$ and CF0 (see \S\ref{sec:S2.2} for definition) on $L_{\mathrm{bol}}$ and $M_{\mathrm{BH}}$, respectively.
We see clear anti-correlations between CF0 and both $L_{\mathrm{bol}}$ and $M_{\mathrm{BH}}$,
consistent with previous studies \citep[e.g.][]{ma_covering_2013, mor_hot-dust_2011, calderone_wise_2012}. %
$\mathrm{EW}$ appears positively correlate with $M_{\mathrm{BH}}$, but we find no apparent correlation between $\mathrm{EW}_{\mathrm{[OIII]5007}}$ and $L_{\mathrm{bol}}$.
This seems against the well known Baldwin effect {
\cite[e.g.][]{baldwin_1977, baldwin_1978, Croom2002, Netzer2004, Kovacevic2010, zhang_baldwin_oiii_2011, Zhang2013}.
This is because 1) the $L_{\mathrm{bol}}$ (= 9.26 * $L_{5100}$) adopted in Fig. \ref{fig:cors} was obtained through fitting the photometric data points, independent to SDSS spectra where [OIII] 5007 EWs were measured, 2) no intrinsic extinction correction was applied to the spectroscopic $L_{5100}$ from \cite{shen_catalog_2011}, and 3) our sample only contains the most luminous sources spanning a rather narrow luminosity range.
Simply adopting the spectroscopic $L_{5100}$ from \cite{shen_catalog_2011} for the whole SDSS DR7 catalog, we do see clear Baldwin effect of [OIII] 5007, with EW $\sim$ $L_{5100}^{-0.21\pm0.009} (r = -0.183)$, consistent with the results reported in literature \citep[e.g.][]{Croom2002,Kovacevic2010,Zhang2013}.
Replacing the spectroscopic $L_{5100}$ from \cite{shen_catalog_2011} with our SED-fitting based $L_{5100}$ (before extinction correction), the derived correlation slope changes to -0.146$\pm$0.007 ($r$ = -0.154) for DR7 quasars. This is because continuum variation (while [OIII] is non-variable) could yield artificial Baldwin effect, since individual quasars in brighter states (thus larger luminosity) tend to have smaller line EW, and vice versa (see \citealt{Jiang2006, Shu2012}).
Comparing line EW with continuum luminosity measured at epochs different from when the spectra were taken could reduce the effect of such bias.
Using extinction-corrected SED-fitting based $L_{5100}$ we further find a slope of -0.088$\pm$0.010 ($r$ = -0.072) for all DR7 quasars, and -0.034$\pm$0.059 ($r$ = -0.015) for the most luminous quasars analyzed in this work.
Clearly, extinction, which could yield lower continuum luminosity and consequently larger [O III] EW, also plays a significant role in the observed Baldwin effect of [O III].
Looking further into the Baldwin effect (and of other lines) is however beyond the scope of this work.
}

We then perform partial correlation to control the effects of bolometric luminosity and SMBH mass, to derive the intrinsic correlations coefficients between $\mathrm{EW}_{\mathrm{[OIII]5007}}$ and CF($\lambda$)
(see Fig. \ref{fig:curve})\footnote{{ Since [O III] EW is approximately $\sim$ $L_{\rm OIII}/L_{\rm bol}$, CF($\lambda$) = $L_\mathrm{IR}$($\lambda$)/$L_{\mathrm{bol}}$, and the effect of $L_{\mathrm{bol}}$ has been controlled in partial correlation analyses, the partial correlation between $L_{\rm OIII}$ and $L_\mathrm{IR}$($\lambda$) is rather similar to that between $\mathrm{EW}_{\mathrm{[OIII]5007}}$ and CF($\lambda$). }}. The  partial correlation coefficients are slightly different from but similar to those of direct correlations. Below we proceed with the partial correlation results. 
The correlation coefficients $r$($\lambda$) derived using CF($\lambda$) obtained with spline fitting and linear interpolating the photometric SED are consistent within statistical uncertainties. 
Therefore, both approaches (spline and linear interpolation), without the need to model the SED with physical templates, could yield consistent scientific results presented below. Hereafter, unless otherwise stated, we adopt the $r$($\lambda$) resulted from linear interpolation (blue line in the lower panel of Fig. \ref{fig:curve}) for further physical interpretation.

We see clear intrinsic  anti-correlations between $\mathrm{EW}_{\mathrm{[OIII]5007}}$ and CF($\lambda$) at short wavelengths. 
The $r$($\lambda$)
reaches a minimum of -0.197 (spline) and -0.195 (linear interpolation) at rest-frame wavelength 2.1 $\micron$, corresponding to p-value of 6.3 $\times$ 10$^{-14}$ and 2.7 $\times$ 10$^{-14}$ for the anti-correlation. 
Therefore, for the first time we have detected statistically significant anti-correlation between $\mathrm{EW}_{\mathrm{[OIII]5007}}$ and dust emission (the simple prediction of of the unification model). %
The $r$($\lambda$) exhibits significant dependence on $\lambda$, %
rising with increasing $\lambda$ at $>$ 2.1 $\micron$.
Note the correlation coefficients also rises with decreasing $\lambda$ at $<$ 2.1 $\micron$. This is because that the significant contribution of the disc component to infrared emission at such short wavelengths could smear out the negative correlation (see Appendix for simulations). Below, while physically interpreting $r$($\lambda$), we focus only on the wavelength range of $\lambda$ $>$ 2.1 \micron.

{ We stress that, though the partial correlation is derived between [O III] EW and $L_\mathrm{IR}$($\lambda$)/$L_{\mathrm{bol}}$ without priorly decomposing the various dust components, the obtained correlation $r$($\lambda$) could be used to unique probe the dust emission in luminous quasars (see \S\ref{sec:discussion}).}
Note since $r$($\lambda$)
was derived based on $L_\mathrm{IR}$($\lambda$) obtained through interpolating the broadband photometric data points, any potential sharp spectral features in $r$($\lambda$) would have been smeared out. In the future if infrared spectra are available for a large sample of sources, we would be able to obtain $r$($\lambda$) with much better spectral resolution to reveal the potential spectral features in it. 

\begin{figure}
    \centering
    \includegraphics[width=0.5\textwidth]{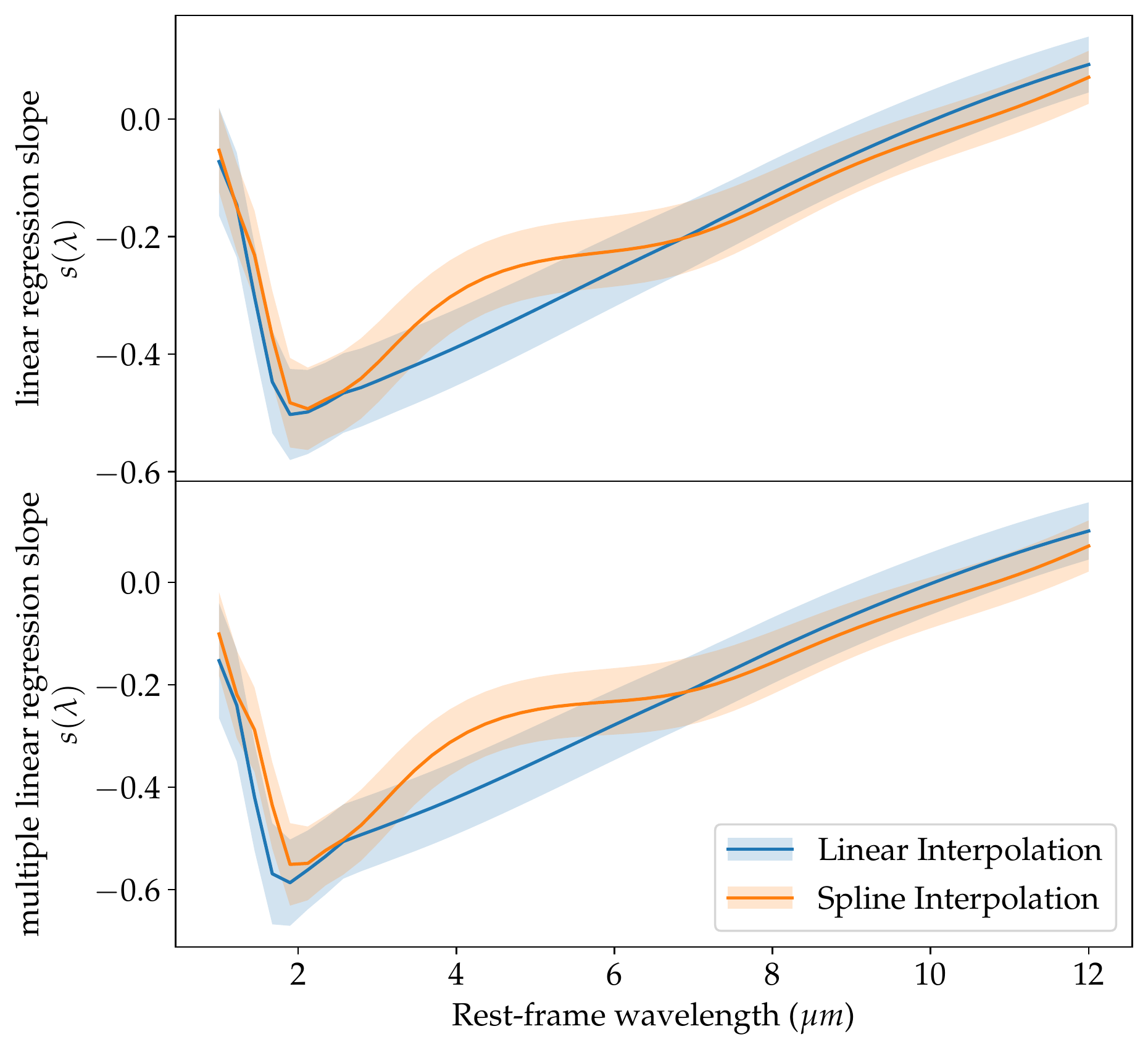}
    \caption{ Similar to the middle and lower panels of Fig. \ref{fig:curve}, but here we plot the linear regression slope (upper panel), and multiple parameter linear regression slope (lower panel, controlling the effects of $L_{\mathrm{bol}}$ and $M$).
    }
    \label{fig:regression}
\end{figure}

{ Finally, we perform linear regression between $\mathrm{EW}_{\mathrm{[OIII]5007}}$ and $L_\mathrm{IR}$($\lambda$)/$L_{\mathrm{bol}}$ (EW $\sim$ [$L_\mathrm{IR}$($\lambda$)/$L_{\mathrm{bol}}]^{s(\lambda)}$), and multiple parameter linear regression (to control the effects of $L_{\mathrm{bol}}$ and SMBH mass, EW $\sim$ [$L_\mathrm{IR}(\lambda)/L_{\mathrm{bol}}]^{s(\lambda)}$*$L_{\mathrm{bol}}^a*M^b)$. 
The resulted regression slope (Fig. \ref{fig:regression}) exhibits statistical significance and $\lambda$ dependence the same as the Pearson correlation coefficient $r(\lambda)$ does, since two approaches (linear regression and Pearson correlation) equivalently quantify the correlation. 
}

\begin{figure*}
    \centering
    \includegraphics[width=\textwidth]{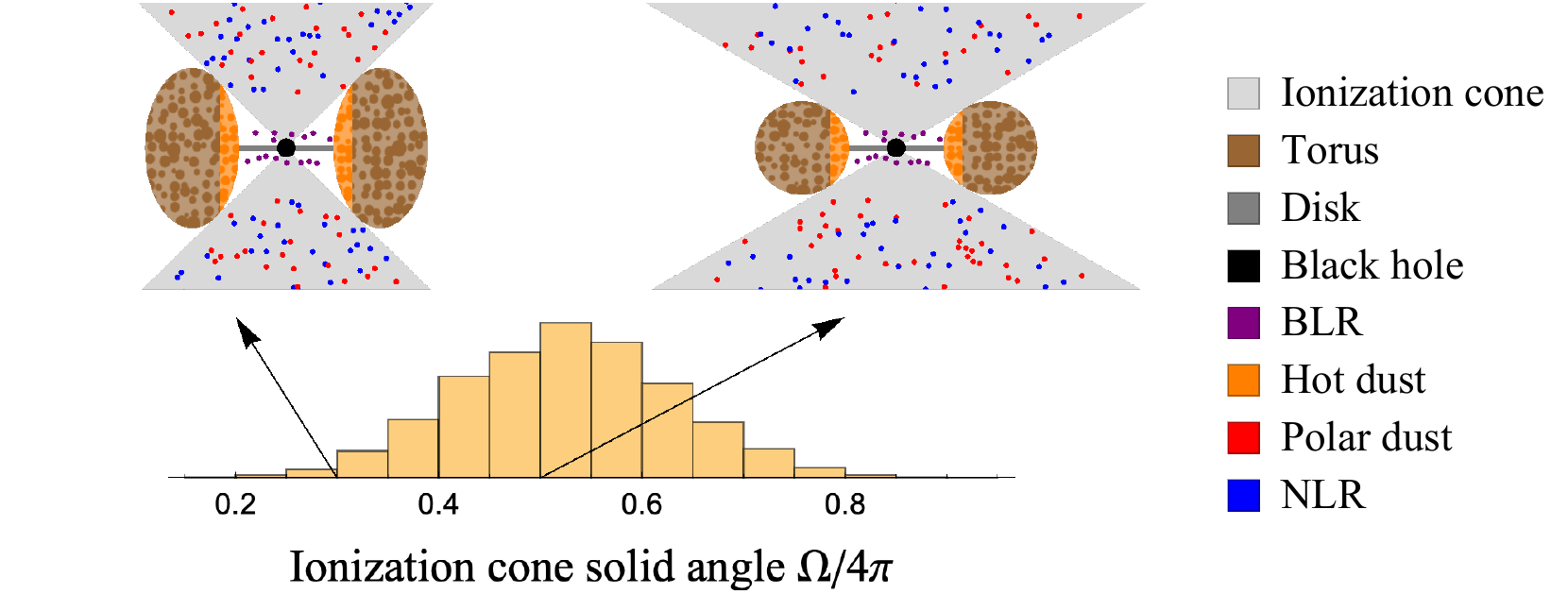}
    \caption{Illustrations of the central structures (color-labelled) of quasars with ionization cone solid angle $\Omega/4\pi=0.3$ and $0.5$, respectively. Larger $\Omega$ would lead to stronger [OIII] 5007 line from NLR and stronger IR emission from the polar dust, both located within the ionization cone, but weaker IR emission from the torus (including hot dust). Assuming a distribution of $\Omega/4\pi$ (yellow histogram, with a mean of 0.5), we could simulate the correlation between [O III] EW and dust emission for a sample of quasars. Figures are not to scale. Note the torus/hot dust are believed to be clumpy \citep{krolik_clumpy_1988,nenkova_agn_2008}, and the simple illustrations here do not imply the torus has smooth boundaries.
    }
    \label{fig:illus1}
\end{figure*}

\section{Discussion}\label{sec:discussion}

\subsection{Simulating the $r$($\lambda$) with a physical model}\label{sec:physicalmodel}

To quantitatively interpret Fig. \ref{fig:curve}, we propose a model as illustrated in Fig. \ref{fig:illus1} based on the unification scheme and known knowledge in the field.
The key gradients in the model include: the central SMBH, accretion disk and BLR, a equatorial torus with the innermost part differently colored as host dust, the bi-polar ionization cones resulted by the torus, and the NLR and polar dust within the ionization cones. 
{ Note in the diagram for simplicity we mark NLR clouds and polar dust clouds separately, however the NLR clouds themselves could be dusty \citep{Netzer1993,Mor2009, Mor2012}, though it is yet unclear whether they dominate the polar dust emission \citep[e.g.][]{Alonso-Herrero2021}.
Another note is that the polar dust could be clumpy like the dusty NLR \citep[e.g.][]{Netzer1993, Groves2006}, thus its intrinsic covering factor could be significant smaller than that of the the polar cone. Nevertheless, we expect its covering factor positively correlates with that of the polar cone.}
 
We then perform simulations to explore whether an intrinsic scatter in the solid angle of the torus (and subsequently of the ionization cone) in a sample of sources could reproduce the observed correlation between $\mathrm{EW}_{\mathrm{[OIII]5007}}$ and $L_{\rm IR}(\lambda)/L_{\rm bol}$. 
{ Below we present analyses to match the derived correlation coefficient $r(\lambda)$ (Fig. \ref{fig:curve}). Utlizing the measured regression slope $s(\lambda)$ and its uncertainty (Fig. \ref{fig:regression}) would yield consistent results.}

In additional to the solid angle of the ionization cone and the equatorial torus, there are other factors which may affect the correlation between $\mathrm{EW}_{\mathrm{[OIII]5007}}$ and CF($\lambda$), and also the large scatter in themselves.
First of all, the UV/optical emission from quasars are significantly variable at timescales of months to years \cite[e.g.][]{Ulrich1997}, while [OIII] and dust emission are produced at larger scales thus are less variable. The variation of the disc component in individual quasars could lead to artificial positive correlation between $\mathrm{EW}_{\mathrm{[OIII]5007}}$ and CF($\lambda$) as both quantities anti-correlate with UV/optical continuum luminosity.
Dust attenuation to UV/optical continuum are significant in some quasars. The attenuation to the continuum, which could lead to larger $\mathrm{EW}_{\mathrm{[OIII]5007}}$, is expected to be stronger in sources with higher solid angle of dust, thus higher CF($\lambda$). 
A similar effect is also expected if the UV SED of quasars has intrinsic scatter. For instance, a harder UV SED could yield larger $\mathrm{EW}_{\mathrm{[OIII]5007}}$ because of the relatively stronger ionization continuum, and meantime stronger re-processed dust emission. Such effects, if significant, would yield $\lambda$-independent positive correlation, contrary to the negative correlation discovered in this work, and shall strengthen the results of this work. Nevertheless, these factors should also be taken into account in our modeling. 

\subsubsection{The anti-correlation between $\mathrm{EW}_{\mathrm{[OIII]5007}}$ and dust emission}\label{S4anticorrelation}

We first assume all dust emission comes from the equatorial torus. 
As the first step, we simulate a set of the ionization cone solid angle $\mu_i$  (normalized by 4$\pi$) randomly sampled from a Gaussian distribution (with a mean of 0.5 for luminous AGNs, \citealt{netzer_revisiting_2015}) and an undetermined standard deviation $\sigma$. We drop nonphysical values of $\mu_i$ ($>$1 or $<$ 0). We further produce a uniform random number between 0 and 1 for each $\mu_i$ and drop $\mu_i$ if it is smaller than the corresponding random number. This is to mimic the observational effect that type 1 quasars are more likely to be detected in systems with larger $\mu_i$.
We then model the $\mathrm{EW}_{\mathrm{[OIII]5007}}$ and torus CF ($L_{\rm IR}(\lambda)/L_{\rm bol}$) as EW$_i$*A$_i$*$\mu_i$ and CF$_i$*A$_i$*(1-$\mu_i$) respectively. Here A$_i$ refers to factors which could simultaneously affect both [OIII] and torus emission per unit solid angle in individual sources, but vary from source to source. An example of such factors is flux variability in individual sources which could yield positive correlation between [OIII] EW and $L_{\rm IR}/L_{\rm bol}$. EW$_i$ and CF$_i$ represent mutually independent factors which affect the emissivity of [OIII] and dust, respectively. Note here we assume the IR emission of the torus linearly correlate with its covering factor and $L_{\rm bol}$ is isotropic (but see \S\ref{sec:anisotropy} for further discussion). 

\begin{figure*}
    \centering
    \includegraphics[width=\textwidth]{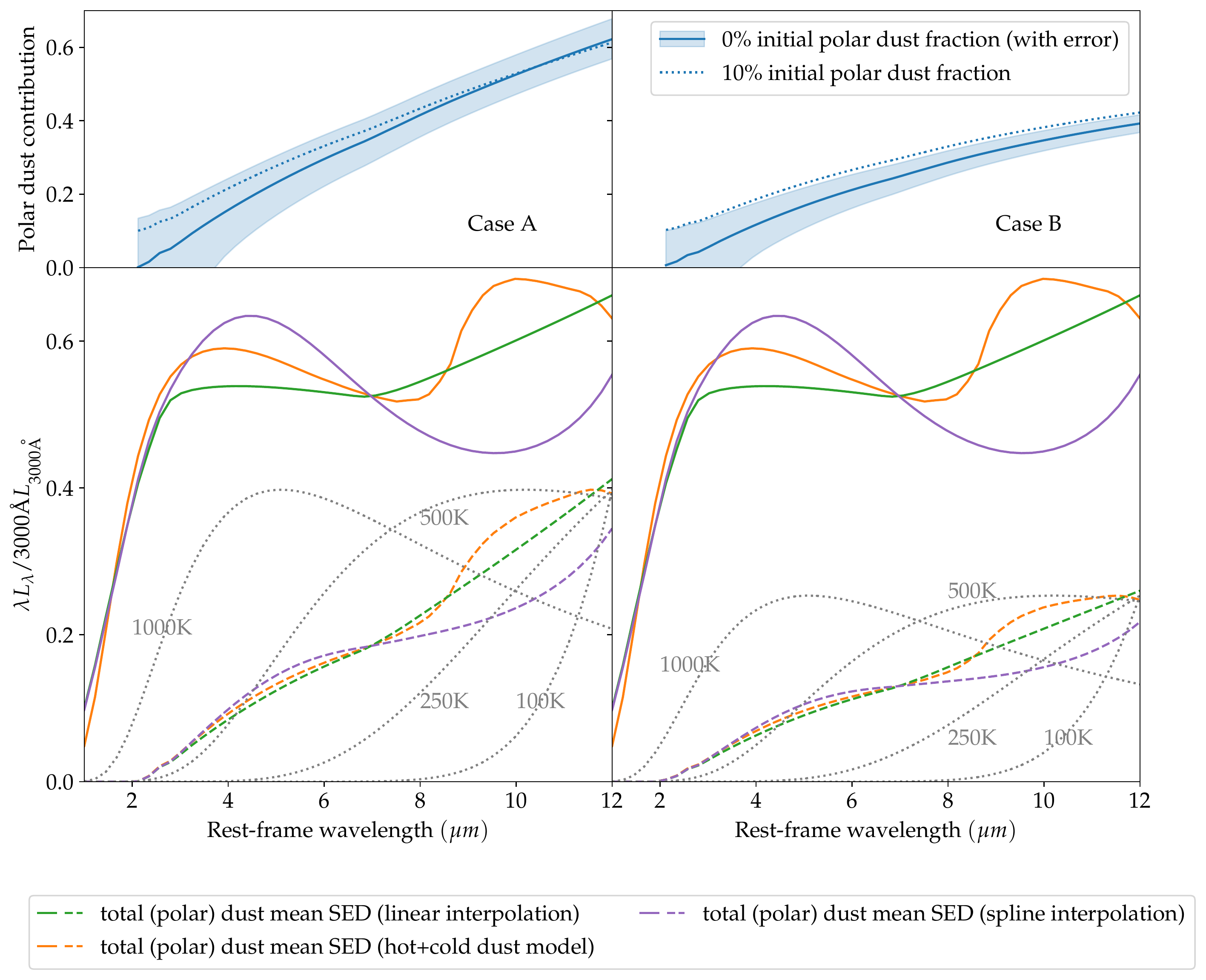}
    \caption{
        Upper: mean polar dust contribution fraction to total dust emission as a function of $\lambda$, converted from $r$($\lambda$) under two extreme conditions (left: Case A; right Case B. see text for definition). The blue lines are derived assuming zero polar dust contribution at 2.1 $\micron$ where the minimum $r$($\lambda$) is detected, and shadows are converted from the statistical uncertainties in $r$($\lambda$) as shown in Fig. \ref{fig:curve}. The dashed lines are derived assuming 10\% polar dust contribution at 2.1$ \micron$.  Lower: The mean SED of the total dust emission (solid lines) of the sample, and of the polar dust emission (dashed lines).
       Gray bodies with temperature of 100K, 250K, 500K and 1000K (dotted) are plotted for comparison. 
        }
    \label{fig:polar}
\end{figure*}

We then consider two extreme situations, and the real situation should be between the two extremes. In Case A: we drop the phase A$_i$, and build a set of EW$_i$ and CF$_i$, both (but mutually independent) following Gaussian distributions in logarithm space with undetermined standard deviations. We then obtain
EW$_i$*$\mu_i$ and CF$_i$*(1-$\mu_i$) to simulate the dependence of $\mathrm{EW}_{\mathrm{[OIII]5007}}$ and torus CF on the solid angle. 
We now calculate the variance and covariance of the simulated EW$_i$*$\mu_i$ and CF$_i$*(1-$\mu_i$), and match the three quantities to those from the real sample (at the wavelength with minimum $r$($\lambda$)) to obtain the three undetermined parameters aforementioned, and then fix them for further studies. 
In Case B, we drop the phase CF$_i$ assuming there in no independent factor which only impacts the dust emissivity per unit solid angle without affecting [OIII] emissivity simultaneously, and the $\mathrm{EW}_{\mathrm{[OIII]5007}}$ and torus CF are modeled as EW$_i$*A$_i$*$\mu_i$ and A$_i$*(1-$\mu_i$). Here the phase EW$_i$ has to be kept, as the observed scatter in $\mathrm{EW}_{\mathrm{[OIII]5007}}$ (0.397 dex) is larger than that of CF0 (0.218 dex) thus there must be additional independent factors which affect only the $\mathrm{EW}_{\mathrm{[OIII]5007}}$.
Following the processes described for Case A, we would also derive the undetermined parameters for Case B. 

To reproduce the minimum $r(\lambda)$ = -0.195 we derived, 
in Case A, we obtain a $\sigma_\mu$ of 0.139, and the scatter of $\mu_i$ could account for 10\% and 52\% of the observed scatter in $\mathrm{EW}_{\mathrm{[OIII]5007}}$ and CF0, respectively. In Case B, a larger $\sigma_\mu$ (0.158) is needed to reproduce the observed anti-correlation. In Case B, $\sigma_\mu$ (and $\sigma_{A_i}$ which is 0.087 dex) account for 13\% and 82\% (5\% and 18\%) of the observed scatter in $\mathrm{EW}_{\mathrm{[OIII]5007}}$ and CF0, respectively.

We stress that in our modeling we have modeled not only the co-variance between $\mathrm{EW}_{\mathrm{[OIII]5007}}$ and CF($\lambda$) but also the variance of themselves. Besides the torus's solid angle, we also have considered the effects of other factors which may affect $\mathrm{EW}_{\mathrm{[OIII]5007}}$ or CF($\lambda$) independently or affect both simultaneously, under two extreme cases (Case A and Case B).   
As shown previously, these additional factors may play dominant roles in producing the observed large scatter in $\mathrm{EW}_{\mathrm{[OIII]5007}}$. Though the variation in the torus's covering factor may only contribute a small fraction ($\sim$ 10-13\%) of the observed scatter of $\mathrm{EW}_{\mathrm{[OIII]5007}}$, thanks to the sufficiently large sample adopted in this work, we have been able to statistically detect its effect, i.e., the anti-correlation between $\mathrm{EW}_{\mathrm{[OIII]5007}}$ and CF($\lambda$). %

\subsubsection{The polar dust contribution}\label{S4polar}

We then simulate the contribution of the polar dust which is assumed to linearly correlate with $\mu_i$.
Now in Case A, the total simulated dust CF is CF$_i$*(1-$\mu_i$) + f($\lambda$)*CF$_i$*$\mu_i$, while $\mathrm{EW}_{\mathrm{[OIII]5007}}$ is modeled as EW$_i$*$\mu_i$. In case B,  we model the total dust CF and  $\mathrm{EW}_{\mathrm{[OIII]5007}}$ as A$_i$*(1-$\mu_i$) + f($\lambda$)*A$_i$*$\mu_i$ and EW$_i$*A$_i$*$\mu_i$, respectively. 
The total dust CF is $\lambda$ dependent, with the second part accounting for emission from the polar dust, and f($\lambda$) the emissivity of polar dust relative to that of torus. 
Since the polar dust emission positively correlates with [OIII] EW, the larger f($\lambda$), the weaker anti-correlation or stronger positive correlation between [OIII] EW and dust CF would be produced. 
Assuming there is zero polar dust contribution at wavelength with minimum $r$($\lambda$), 
and matching the covariance between simulated EW$_i$*$\mu_i$ and CF$_i$*(1-$\mu_i$) + f($\lambda$)*CF$_i$*$\mu_i$ with the observed $r$($\lambda$), we could solve out f($\lambda$) and then calculate the average fraction of polar dust contribution to total dust emission (as shown in Fig. \ref{fig:polar}). 
Note in Case B weaker polar dust contribution is required at given $\lambda$ compared with Case A. This is because in Case A, only the term $\mu_i$ contributes to the positive correlation between [OIII] and polar dust emission, while in Case B both $\mu_i$ and A$_i$ make the contribution.

The polar dust contribution fraction
rises from 0 at 2.1 $\micron$ (as assumed) to between $\sim$ 39\% and 62\% (for Case B and Case A respectively) at rest-frame $12~\micron$, and overall contribution of the polar dust from $1 - 12~\micron$ is between $\sim$ 16\%--23\%.
In Fig. \ref{fig:polar} we plot the mean dust SEDs for our sample. The mean SEDs derived from various fitting/iterpolating approaches are somehow different. For spline fitting and linear interpolation, we subtract the best-fit disc component obtained with the hot+cold dust model SED fitting, to derive the dust component. Note none of these mean dust SEDs is perfect, considering the hot+cold dust model fitting yielded systematical residuals (see Fig. \ref{fig:residual}), while spline and linear interpolation have smeared out physical features in the mean SED. 

The mean polar dust SED could then be derived using the polar dust contribution fraction we obtained. 
Gray body of 1000 K, 500 K, 250K and 100 K (with an emissivity index of 1.6, \citealt{toba_how_2021}) are over-plotted in Fig. \ref{fig:polar} for comparison. Clearly the mean polar dust SED we derived is considerably redder than that of the total dust emission. Fitting them approximately with a gray body yields temperature of $\sim$ 405-559 K. Unlike the fraction of polar dust contribution, the shape of the derived mean polar dust SED is less sensitive to the condition we assumed (Case A or B, see Fig. \ref{fig:polar}).
These mean SEDs (available at \url{http://home.ustc.edu.cn/~wul/emil.html}) could be adopted in future SED fittings. 

{ Note due to the limited wavelength range, the approximate gray body temperature we derived could be subject to significant uncertainties.
Nevertheless, 
a redder polar dust emission is qualitatively consistent with the NLR dust models, e.g., \cite{Schweitzer2008} and \cite{Mor2009}, which calculated the expected emission of NLR dust, assuming the NLR dust is concentrated in a thin spherical shell. 
\cite{Mor2009} obtained the mean NLR dust emission of 26 quasars through fitting their Spitzer 2-35$\micron$ specrtra with three dust components (hot dust, torus, and NLR dust). While their results could be significantly model/template dependent, it is interesting to note their best-fit NLR dust emission appears even redder, and their mean NLR dust contribution fraction rapidly drops at $<$ 10 $\micron$, becoming negligible at $<$ 8$\micron$ (Fig. 4 of \citealt{Mor2009}). This suggests that either NLR is not the only location of polar dust that warmer polar dust also exists at scales smaller than NLR, or a thin spherical shell model for NLR dust is over-simplified. The fact that the mean polar dust SED we derived in Fig. \ref{fig:polar} clearly deviates from a single temperature gray body also suggests a complicated geometry/distribution of the polar dust. 
}
Note again any potential sharp spectral features { (such as the silicate emission feature at $\sim$ 10$\micron$ expected from the dusty NLR, \citealt{Schweitzer2008,Mor2012} )} would also have been smeared out in the mean polar dust SED. In the future applying the technique we developed in this work to a large set of infrared spectra, we would be able obtain a mean spectrum (instead of a smoothed SED) of polar dust in quasars.

We note that it is a bit arbitrary to assume zero polar dust contribution at wavelength with minimum $r$($\lambda$). Nevertheless, it has to be rather small at short wavelengths considering the observed negative correlation. In fact the mean polar dust SED derived from modeling $r$($\lambda$) is insensitive to this assumption. For instance, assuming a 10\% contribution of polar dust to total dust emission at wavelength with minimum $r$($\lambda$) barely changes the overall results (see the blue dashed line in the upper panels of Fig. \ref{fig:polar}).

Another note is that we have assumed a mean torus covering factor of 0.5  \citep[][]{netzer_revisiting_2015} in our modeling.
In fact, assuming $L_{\rm dust}/L_{7.8\micron}$ $\sim$ 3 \citep{Weedman2012}, we obtain a median $L_{\rm dust}/L_{\rm bol}$ $\sim$ 0.314 for our sample.  Considering the extra uncertainties involved when converting $L_{\rm dust}/L_{\rm bol}$ to torus covering factor, including the uncertainties in bolometric luminosity, possible sample selection bias\footnote{Note our sample was selected to have $L_{\rm bol}$ (converted from $L_{5100\AA}$) above 10$^{46}$ erg/s. Sources selected to have large $L_{\rm bol}$ naturally tend to have smaller $L_{\rm dust}/L_{\rm bol}$.} and possible anisotropy in central emission \citep{netzer_revisiting_2015}, we have adopted 0.5 instead of 0.314 in this work. Meanwhile, utilizing smaller mean torus covering factor does alter the derived mean polar dust contribution fraction (the smaller mean torus covering factor, the larger mean polar dust contribution fraction, as expected). However the shape of the derived mean polar dust SED barely changes.

\subsection{Effects of possibly anisotropic of disk radiation}\label{sec:anisotropy}

In \S\ref{sec:physicalmodel} we have assumed the dust emission (of torus and polar dust) linearly correlates with the solid angle of torus and the ionization cone, respectively. This assumption is valid if both the dust and disk emission are isotropic. While for type 1 quasars it is reasonable to assume the torus emission is isotropic \citep{stalevski_torus_2012, stalevski_dust_2016}, and so does the polar dust emission at larger scale, 
the radiation of the accretion disk might be anisotropic as proposed by \cite{netzer_1987}:
\begin{equation}\label{equation:1}
L(\theta) \propto \mathrm{cos}\theta (2\mathrm{cos}\theta + 1)
\end{equation}
where $\theta$ is the angle between the line of sight and the axis of the disk.
The anisotropic disk radiation would subsequently yield  a nonlinear relation between $L_{\rm IR}/L_{\rm bol}$ and torus covering factor (see the middle panel of Figure 7 in \citealt{stalevski_dust_2016}).
Nevertheless, because the relation is strictly monotonic, it has little effect on the correlation coefficients we presented in \S\ref{sec:S3}, though it could affect the physical modeling to $r$($\lambda$) in \S\ref{sec:physicalmodel}.

The $\mathrm{EW}_{\mathrm{[OIII]5007}}$ could also be sensitive to the inclination angle if the disk emission is anisotropic \citep{Risaliti2011}.
\cite{Bisogni2019} found type 1 quasars with larger $\mathrm{EW}_{\mathrm{[OIII]5007}}$ tend to have redder infrared SEDs, and attribute it to the inclination effect that the torus emission is redder at larger inclination angle.  However, the inclination angle dependence of the torus emission in type 1 quasars is expected to be rather weak as the clouds are believed to be clumpy \cite[e.g.][]{stalevski_torus_2012}.
Furthermore, in case of anisotropic disk emission, as the observed $L_{\mathrm{UV}}$ would be sensitive to the inclination, and both $L_\mathrm{IR}$($\lambda$)/$L_{\mathrm{bol}}$ and $\mathrm{EW}_{\mathrm{[OIII]5007}}$ anti-correlate with the observed $L_{\mathrm{UV}}$, the inclination effect should yield artificial positive $r$($\lambda$) between $L_\mathrm{IR}$($\lambda$)/$L_{\mathrm{bol}}$ and $\mathrm{EW}_{\mathrm{[OIII]5007}}$, contrary to the clear negative correlation we discovered.
If the disk emission is indeed anisotropic as shown in Equation \ref{equation:1},  
we could develop a fix to the modeling of $r$($\lambda$) given in \S\ref{sec:physicalmodel}.
We previously model $L_\mathrm{IR}$ %
as CF$_i$*A$_i$*(1-$\mu_i$), and it now needs to be corrected to CF$_i$*A$_i$*k($\mu_i$)(1-$\mu_i$), where k($\mu_i$) is the nonlinear correction function from torus covering factor to $L_{\rm IR}$ given by \cite{stalevski_dust_2016}. 
Compared with \S\ref{sec:physicalmodel}, a considerably larger scatter in $\mu_i$ would be then required to explain the observed negative $r$($\lambda$) at short wavelength.
Furthermore, including the factor k($\mu_i$) would yield significantly larger scatter in $L_{\rm IR}$ at given scatter of $\mu_i$ (see Figure 7 in \citealt{stalevski_dust_2016}).  
Taking these new factors into consideration, 
we failed to reproduce the observed $r$($\lambda$) following the procedures in \S\ref{sec:physicalmodel}, as the observed scatter of  $L_\mathrm{IR}$/$L_{\mathrm{bol}}$ (0.218 dex for CF0) is too small to enable sufficiently large scatter in $\mu_i$. 
Thus the anisotropic scenario of disk emission is disfavored.
While anisotropy weaker than Equation \ref{equation:1} could still be feasible, %
we would defer further exploration on this issue to a future dedicated work.

We note the $\mathrm{EW}_{\mathrm{[OIII]5007}}$ dependent infrared SED slope reported by \cite{Bisogni2019} may qualitatively explained by our model that quasars with higher solid angle of the ionization cone (thus larger $\mathrm{EW}_{\mathrm{[OIII]5007}}$) have stronger contribution from the cooler polar dust (thus redder infrared SED). Quantitative analyses are deferred to a future work.
Interestingly, a most recent study of \cite{Lyu2022} also found quasars with larger forbidden line EWs tend to have redder infrared colors, and they attributed this correlation to stronger polar dust contribution (which is well consistent with the scenario developed in this work).

\section{Summary}

We analyze the correlation between [OIII] 5007 EW and near- to mid- infrared dust emission of a sample of luminous SDSS quasars with bolometric luminosity above $10^{46}$ erg/s. The ratio of infrared to bolometric luminosity  $L_\mathrm{IR}$($\lambda$)/$L_{\mathrm{bol}}$ for each quasar is measured through interpolating the photometric data from SDSS, 2MASS and WISE.
The correlation efficient $r$($\lambda$) between $\mathrm{EW}_{\mathrm{[OIII]5007}}$ and $L_\mathrm{IR}$($\lambda$)/$L_{\mathrm{bol}}$ for the sample is obtained as a function of infrared wavelength $\lambda$. We derive the following key findings:\\

1. We detect statistically robust anti-correlation between  $\mathrm{EW}_{\mathrm{[OIII]5007}}$ and hot dust emission in luminous quasars. This directly confirms the ``simple but had never been confirmed" prediction of the unification scheme of AGNs, that sources with larger solid angle of the equatorial torus (thus stronger dust emission) must have smaller solid angle of the ionization cone (thus weaker [OIII] emission).\\

2. The anti-correlation between $\mathrm{EW}_{\mathrm{[OIII]5007}}$ and $L_\mathrm{IR}$($\lambda$)/$L_{\mathrm{bol}}$ gradually weakens toward longer wavelength 
(at which $L_\mathrm{IR}$($\lambda$)/$L_{\mathrm{bol}}$ is calculated) 
from 2 to 12 $\micron$. 
This indicates there exists an extra cooler dust component, i.e., the polar dust, with its emission positively correlating with the solid angle of the ionization cone, which could reduce the anti-correlation between  $\mathrm{EW}_{\mathrm{[OIII]5007}}$) and total dust emission at longer wavelengths. \\

3. Fitting the derived anti-correlation efficient $r$($\lambda$) with a physical model based on the unification scheme, we find the polar dust contribution to total dust emission gradually increases with wavelength, and reaches between 39\%--62\% (model dependent) at rest frame 12\micron. Multiplying the mean total dust SED of the sample and the polar dust contribution fraction we derived, we obtain a mean polar dust SED which could be used for future SED fitting. \\

\section*{Acknowledgements}

{ We thank the anonymous referee for constructive comments
that have significantly improved the manuscript.}
This work was supported by the National Science Foundation of China (No. 1890693, 12033006 $\&$ 12192221). The authors gratefully acknowledge the support of Cyrus Chun Ying Tang Foundations.

\subsection*{Data availability}

All data used in this work comes from the Sloan Digital Sky Survey data release 7 quasar catalogue, which can be downloaded from \url{https://users.obs.carnegiescience.edu/yshen/BH_mass/dr7.htm}.

\bibliographystyle{mnras}
\bibliography{paper}

\begin{thebibliography}{}
\makeatletter
\relax
\def\mn@urlcharsother{\let\do\@makeother \do\$\do\&\do\#\do\^\do\_\do\%\do\~}
\def\mn@doi{\begingroup\mn@urlcharsother \@ifnextchar [ {\mn@doi@}
  {\mn@doi@[]}}
\def\mn@doi@[#1]#2{\def\@tempa{#1}\ifx\@tempa\@empty \href
  {http://dx.doi.org/#2} {doi:#2}\else \href {http://dx.doi.org/#2} {#1}\fi
  \endgroup}
\def\mn@eprint#1#2{\mn@eprint@#1:#2::\@nil}
\def\mn@eprint@arXiv#1{\href {http://arxiv.org/abs/#1} {{\tt arXiv:#1}}}
\def\mn@eprint@dblp#1{\href {http://dblp.uni-trier.de/rec/bibtex/#1.xml}
  {dblp:#1}}
\def\mn@eprint@#1:#2:#3:#4\@nil{\def\@tempa {#1}\def\@tempb {#2}\def\@tempc
  {#3}\ifx \@tempc \@empty \let \@tempc \@tempb \let \@tempb \@tempa \fi \ifx
  \@tempb \@empty \def\@tempb {arXiv}\fi \@ifundefined
  {mn@eprint@\@tempb}{\@tempb:\@tempc}{\expandafter \expandafter \csname
  mn@eprint@\@tempb\endcsname \expandafter{\@tempc}}}

\bibitem[\protect\citeauthoryear{{Alonso-Herrero} et~al.,}{{Alonso-Herrero}
  et~al.}{2021}]{Alonso-Herrero2021}
{Alonso-Herrero} A.,  et~al., 2021, \mn@doi [\aap]
  {10.1051/0004-6361/202141219}, \href
  {https://ui.adsabs.harvard.edu/abs/2021A&A...652A..99A} {652, A99}

\bibitem[\protect\citeauthoryear{Antonucci}{Antonucci}{1993}]{antonucci_unified_1993}
Antonucci R.,  1993, \mn@doi [Annual Review of Astronomy and Astrophysics]
  {10.1146/annurev.aa.31.090193.002353}, 31, 473

\bibitem[\protect\citeauthoryear{Asmus}{Asmus}{2019}]{asmus_new_2019}
Asmus D.,  2019, \mn@doi [Monthly Notices of the Royal Astronomical Society]
  {10.1093/mnras/stz2289}, 489, 2177

\bibitem[\protect\citeauthoryear{{Asmus}, {H{\"o}nig}  \& {Gandhi}}{{Asmus}
  et~al.}{2016}]{Asmus2016}
{Asmus} D.,  {H{\"o}nig} S.~F.,   {Gandhi} P.,  2016, \mn@doi [\apj]
  {10.3847/0004-637X/822/2/109}, \href
  {https://ui.adsabs.harvard.edu/abs/2016ApJ...822..109A} {822, 109}

\bibitem[\protect\citeauthoryear{{Baldwin}}{{Baldwin}}{1977}]{baldwin_1977}
{Baldwin} J.~A.,  1977, \mn@doi [\apj] {10.1086/155294}, \href
  {https://ui.adsabs.harvard.edu/abs/1977ApJ...214..679B} {214, 679}

\bibitem[\protect\citeauthoryear{{Baldwin}, {Burke}, {Gaskell}  \&
  {Wampler}}{{Baldwin} et~al.}{1978}]{baldwin_1978}
{Baldwin} J.~A.,  {Burke} W.~L.,  {Gaskell} C.~M.,   {Wampler} E.~J.,  1978,
  \mn@doi [\nat] {10.1038/273431a0}, \href
  {https://ui.adsabs.harvard.edu/abs/1978Natur.273..431B} {273, 431}

\bibitem[\protect\citeauthoryear{{Barvainis}}{{Barvainis}}{1987}]{barvainis_hot_dust_1987}
{Barvainis} R.,  1987, \mn@doi [\apj] {10.1086/165571}, \href
  {https://ui.adsabs.harvard.edu/abs/1987ApJ...320..537B} {320, 537}

\bibitem[\protect\citeauthoryear{{Bisogni}, {Lusso}, {Marconi}  \&
  {Risaliti}}{{Bisogni} et~al.}{2019}]{Bisogni2019}
{Bisogni} S.,  {Lusso} E.,  {Marconi} A.,   {Risaliti} G.,  2019, \mn@doi
  [\mnras] {10.1093/mnras/stz495}, \href
  {https://ui.adsabs.harvard.edu/abs/2019MNRAS.485.1405B} {485, 1405}

\bibitem[\protect\citeauthoryear{{Calderone}, {Sbarrato}  \&
  {Ghisellini}}{{Calderone} et~al.}{2012}]{calderone_wise_2012}
{Calderone} G.,  {Sbarrato} T.,   {Ghisellini} G.,  2012, \mn@doi [\mnras]
  {10.1111/j.1745-3933.2012.01296.x}, \href
  {https://ui.adsabs.harvard.edu/abs/2012MNRAS.425L..41C} {425, L41}

\bibitem[\protect\citeauthoryear{Carpenter}{Carpenter}{2001}]{carpenter_color_2001}
Carpenter J.~M.,  2001, \mn@doi [The Astronomical Journal] {10.1086/320383},
  121, 2851

\bibitem[\protect\citeauthoryear{{Croom} et~al.,}{{Croom}
  et~al.}{2002}]{Croom2002}
{Croom} S.~M.,  et~al., 2002, \mn@doi [\mnras]
  {10.1046/j.1365-8711.2002.05910.x}, \href
  {https://ui.adsabs.harvard.edu/abs/2002MNRAS.337..275C} {337, 275}

\bibitem[\protect\citeauthoryear{Eaton}{Eaton}{1983}]{eaton_multivariate_1983}
Eaton M.,  1983, Multivariate {Statistics}: {A} {Vector} {Space} {Approach}.
Wiley, pp 116--117, \url {https://books.google.com.au/books?id=1CvvAAAAMAAJ}

\bibitem[\protect\citeauthoryear{{Eisenstein} et~al.,}{{Eisenstein}
  et~al.}{2011}]{sdss}
{Eisenstein} D.~J.,  et~al., 2011, \mn@doi [\aj] {10.1088/0004-6256/142/3/72},
  \href {https://ui.adsabs.harvard.edu/abs/2011AJ....142...72E} {142, 72}

\bibitem[\protect\citeauthoryear{{Evans}, {Ford}, {Kinney}, {Antonucci},
  {Armus}  \& {Caganoff}}{{Evans} et~al.}{1991}]{Evans1991}
{Evans} I.~N.,  {Ford} H.~C.,  {Kinney} A.~L.,  {Antonucci} R.~R.~J.,  {Armus}
  L.,   {Caganoff} S.,  1991, \mn@doi [\apjl] {10.1086/185951}, \href
  {https://ui.adsabs.harvard.edu/abs/1991ApJ...369L..27E} {369, L27}

\bibitem[\protect\citeauthoryear{Fukugita, Ichikawa, Gunn, Doi, Shimasaku  \&
  Schneider}{Fukugita et~al.}{1996}]{fukugita_sloan_1996}
Fukugita M.,  Ichikawa T.,  Gunn J.~E.,  Doi M.,  Shimasaku K.,   Schneider
  D.~P.,  1996, \mn@doi [The Astronomical Journal] {10.1086/117915}, 111, 1748

\bibitem[\protect\citeauthoryear{Gordon, Clayton, Misselt, Landolt  \&
  Wolff}{Gordon et~al.}{2003}]{gordon_quantitative_2003}
Gordon K.~D.,  Clayton G.~C.,  Misselt K.~A.,  Landolt A.~U.,   Wolff M.~J.,
  2003, \mn@doi [The Astrophysical Journal] {10.1086/376774}, 594, 279

\bibitem[\protect\citeauthoryear{{Groves}, {Dopita}  \& {Sutherland}}{{Groves}
  et~al.}{2006}]{Groves2006}
{Groves} B.,  {Dopita} M.,   {Sutherland} R.,  2006, \mn@doi [\aap]
  {10.1051/0004-6361:20065097}, \href
  {https://ui.adsabs.harvard.edu/abs/2006A&A...458..405G} {458, 405}

\bibitem[\protect\citeauthoryear{{H{\"o}nig} \& {Kishimoto}}{{H{\"o}nig} \&
  {Kishimoto}}{2017}]{Honig2017}
{H{\"o}nig} S.~F.,  {Kishimoto} M.,  2017, \mn@doi [\apjl]
  {10.3847/2041-8213/aa6838}, \href
  {https://ui.adsabs.harvard.edu/abs/2017ApJ...838L..20H} {838, L20}

\bibitem[\protect\citeauthoryear{{Jaffe}, {Ford}, {Ferrarese}, {van den Bosch}
  \& {O'Connell}}{{Jaffe} et~al.}{1996}]{Jaffe1996}
{Jaffe} W.,  {Ford} H.,  {Ferrarese} L.,  {van den Bosch} F.,   {O'Connell}
  R.~W.,  1996, \mn@doi [\apj] {10.1086/176963}, \href
  {https://ui.adsabs.harvard.edu/abs/1996ApJ...460..214J} {460, 214}

\bibitem[\protect\citeauthoryear{{Jiang}, {Wang}  \& {Wang}}{{Jiang}
  et~al.}{2006}]{Jiang2006}
{Jiang} P.,  {Wang} J.~X.,   {Wang} T.~G.,  2006, \mn@doi [\apj]
  {10.1086/503866}, \href
  {https://ui.adsabs.harvard.edu/abs/2006ApJ...644..725J} {644, 725}

\bibitem[\protect\citeauthoryear{{Kellermann}, {Sramek}, {Schmidt}, {Shaffer}
  \& {Green}}{{Kellermann} et~al.}{1989}]{Kellermann1989}
{Kellermann} K.~I.,  {Sramek} R.,  {Schmidt} M.,  {Shaffer} D.~B.,   {Green}
  R.,  1989, \mn@doi [\aj] {10.1086/115207}, \href
  {https://ui.adsabs.harvard.edu/abs/1989AJ.....98.1195K} {98, 1195}

\bibitem[\protect\citeauthoryear{{Kova{\v{c}}evi{\'c}}, {Popovi{\'c}}  \&
  {Dimitrijevi{\'c}}}{{Kova{\v{c}}evi{\'c}} et~al.}{2010}]{Kovacevic2010}
{Kova{\v{c}}evi{\'c}} J.,  {Popovi{\'c}} L.~{\v{C}}.,   {Dimitrijevi{\'c}}
  M.~S.,  2010, \mn@doi [\apjs] {10.1088/0067-0049/189/1/15}, \href
  {https://ui.adsabs.harvard.edu/abs/2010ApJS..189...15K} {189, 15}

\bibitem[\protect\citeauthoryear{{Krolik} \& {Begelman}}{{Krolik} \&
  {Begelman}}{1988}]{krolik_clumpy_1988}
{Krolik} J.~H.,  {Begelman} M.~C.,  1988, \mn@doi [\apj] {10.1086/166414},
  \href {https://ui.adsabs.harvard.edu/abs/1988ApJ...329..702K} {329, 702}

\bibitem[\protect\citeauthoryear{{Lawrence} et~al.,}{{Lawrence}
  et~al.}{2007}]{ukidss}
{Lawrence} A.,  et~al., 2007, \mn@doi [\mnras]
  {10.1111/j.1365-2966.2007.12040.x}, \href
  {https://ui.adsabs.harvard.edu/abs/2007MNRAS.379.1599L} {379, 1599}

\bibitem[\protect\citeauthoryear{Leftley, Tristram, Hönig, Asmus, Kishimoto
  \& Gandhi}{Leftley et~al.}{2021}]{leftley_resolving_2021}
Leftley J.~H.,  Tristram K. R.~W.,  Hönig S.~F.,  Asmus D.,  Kishimoto M.,
  Gandhi P.,  2021, \mn@doi [The Astrophysical Journal]
  {10.3847/1538-4357/abee80}, 912, 96

\bibitem[\protect\citeauthoryear{{L{\'o}pez-Gonzaga}, {Burtscher}, {Tristram},
  {Meisenheimer}  \& {Schartmann}}{{L{\'o}pez-Gonzaga}
  et~al.}{2016}]{Lopez2016}
{L{\'o}pez-Gonzaga} N.,  {Burtscher} L.,  {Tristram} K.~R.~W.,  {Meisenheimer}
  K.,   {Schartmann} M.,  2016, \mn@doi [\aap] {10.1051/0004-6361/201527590},
  \href {https://ui.adsabs.harvard.edu/abs/2016A&A...591A..47L} {591, A47}

\bibitem[\protect\citeauthoryear{{Lyke} et~al.,}{{Lyke}
  et~al.}{2020}]{sdss_dr16}
{Lyke} B.~W.,  et~al., 2020, \mn@doi [\apjs] {10.3847/1538-4365/aba623}, \href
  {https://ui.adsabs.harvard.edu/abs/2020ApJS..250....8L} {250, 8}

\bibitem[\protect\citeauthoryear{{Lyu} \& {Rieke}}{{Lyu} \&
  {Rieke}}{2022}]{Lyu2022}
{Lyu} J.,  {Rieke} G.~H.,  2022, arXiv e-prints, \href
  {https://ui.adsabs.harvard.edu/abs/2022arXiv221008037L} {p. arXiv:2210.08037}

\bibitem[\protect\citeauthoryear{Ma \& Wang}{Ma \&
  Wang}{2013}]{ma_covering_2013}
Ma X.-C.,  Wang T.-G.,  2013, \mn@doi [Monthly Notices of the Royal
  Astronomical Society] {10.1093/mnras/stt143}, 430, 3445

\bibitem[\protect\citeauthoryear{{Maiolino}, {Shemmer}, {Imanishi}, {Netzer},
  {Oliva}, {Lutz}  \& {Sturm}}{{Maiolino} et~al.}{2007}]{Maiolino2007}
{Maiolino} R.,  {Shemmer} O.,  {Imanishi} M.,  {Netzer} H.,  {Oliva} E.,
  {Lutz} D.,   {Sturm} E.,  2007, \mn@doi [\aap] {10.1051/0004-6361:20077252},
  \href {https://ui.adsabs.harvard.edu/abs/2007A&A...468..979M} {468, 979}

\bibitem[\protect\citeauthoryear{{Mor} \& {Netzer}}{{Mor} \&
  {Netzer}}{2012}]{Mor2012}
{Mor} R.,  {Netzer} H.,  2012, \mn@doi [\mnras]
  {10.1111/j.1365-2966.2011.20060.x}, \href
  {https://ui.adsabs.harvard.edu/abs/2012MNRAS.420..526M} {420, 526}

\bibitem[\protect\citeauthoryear{{Mor} \& {Trakhtenbrot}}{{Mor} \&
  {Trakhtenbrot}}{2011}]{mor_hot-dust_2011}
{Mor} R.,  {Trakhtenbrot} B.,  2011, \mn@doi [\apjl]
  {10.1088/2041-8205/737/2/L36}, \href
  {https://ui.adsabs.harvard.edu/abs/2011ApJ...737L..36M} {737, L36}

\bibitem[\protect\citeauthoryear{{Mor}, {Netzer}  \& {Elitzur}}{{Mor}
  et~al.}{2009}]{Mor2009}
{Mor} R.,  {Netzer} H.,   {Elitzur} M.,  2009, \mn@doi [\apj]
  {10.1088/0004-637X/705/1/298}, \href
  {https://ui.adsabs.harvard.edu/abs/2009ApJ...705..298M} {705, 298}

\bibitem[\protect\citeauthoryear{{M{\"u}ller-S{\'a}nchez}, {Prieto}, {Hicks},
  {Vives-Arias}, {Davies}, {Malkan}, {Tacconi}  \&
  {Genzel}}{{M{\"u}ller-S{\'a}nchez} et~al.}{2011}]{muller_nlr_2011}
{M{\"u}ller-S{\'a}nchez} F.,  {Prieto} M.~A.,  {Hicks} E.~K.~S.,  {Vives-Arias}
  H.,  {Davies} R.~I.,  {Malkan} M.,  {Tacconi} L.~J.,   {Genzel} R.,  2011,
  \mn@doi [\apj] {10.1088/0004-637X/739/2/69}, \href
  {https://ui.adsabs.harvard.edu/abs/2011ApJ...739...69M} {739, 69}

\bibitem[\protect\citeauthoryear{{Nenkova}, {Sirocky}, {Nikutta}, {Ivezi{\'c}}
  \& {Elitzur}}{{Nenkova} et~al.}{2008}]{nenkova_agn_2008}
{Nenkova} M.,  {Sirocky} M.~M.,  {Nikutta} R.,  {Ivezi{\'c}} {\v{Z}}.,
  {Elitzur} M.,  2008, \mn@doi [\apj] {10.1086/590483}, \href
  {https://ui.adsabs.harvard.edu/abs/2008ApJ...685..160N} {685, 160}

\bibitem[\protect\citeauthoryear{{Netzer}}{{Netzer}}{1987}]{netzer_1987}
{Netzer} H.,  1987, \mn@doi [\mnras] {10.1093/mnras/225.1.55}, \href
  {https://ui.adsabs.harvard.edu/abs/1987MNRAS.225...55N} {225, 55}

\bibitem[\protect\citeauthoryear{Netzer}{Netzer}{2015}]{netzer_revisiting_2015}
Netzer H.,  2015, \mn@doi [Annual Review of Astronomy and Astrophysics]
  {10.1146/annurev-astro-082214-122302}, 53, 365

\bibitem[\protect\citeauthoryear{{Netzer} \& {Laor}}{{Netzer} \&
  {Laor}}{1993}]{Netzer1993}
{Netzer} H.,  {Laor} A.,  1993, \mn@doi [\apjl] {10.1086/186741}, \href
  {https://ui.adsabs.harvard.edu/abs/1993ApJ...404L..51N} {404, L51}

\bibitem[\protect\citeauthoryear{{Netzer}, {Shemmer}, {Maiolino}, {Oliva},
  {Croom}, {Corbett}  \& {di Fabrizio}}{{Netzer} et~al.}{2004}]{Netzer2004}
{Netzer} H.,  {Shemmer} O.,  {Maiolino} R.,  {Oliva} E.,  {Croom} S.,
  {Corbett} E.,   {di Fabrizio} L.,  2004, \mn@doi [\apj] {10.1086/423608},
  \href {https://ui.adsabs.harvard.edu/abs/2004ApJ...614..558N} {614, 558}

\bibitem[\protect\citeauthoryear{{Nikutta}, {Lopez-Rodriguez}, {Ichikawa},
  {Levenson}, {Packham}, {H{\"o}nig}  \& {Alonso-Herrero}}{{Nikutta}
  et~al.}{2021}]{Nikutta2021}
{Nikutta} R.,  {Lopez-Rodriguez} E.,  {Ichikawa} K.,  {Levenson} N.~A.,
  {Packham} C.,  {H{\"o}nig} S.~F.,   {Alonso-Herrero} A.,  2021, \mn@doi
  [\apj] {10.3847/1538-4357/ac06a6}, \href
  {https://ui.adsabs.harvard.edu/abs/2021ApJ...919..136N} {919, 136}

\bibitem[\protect\citeauthoryear{{Risaliti}, {Salvati}  \&
  {Marconi}}{{Risaliti} et~al.}{2011}]{Risaliti2011}
{Risaliti} G.,  {Salvati} M.,   {Marconi} A.,  2011, \mn@doi [\mnras]
  {10.1111/j.1365-2966.2010.17843.x}, \href
  {https://ui.adsabs.harvard.edu/abs/2011MNRAS.411.2223R} {411, 2223}

\bibitem[\protect\citeauthoryear{Roseboom, Lawrence, Elvis, Petty, Shen  \&
  Hao}{Roseboom et~al.}{2013}]{roseboom_ir-derived_2013}
Roseboom I.~G.,  Lawrence A.,  Elvis M.,  Petty S.,  Shen Y.,   Hao H.,  2013,
  \mn@doi [Monthly Notices of the Royal Astronomical Society]
  {10.1093/mnras/sts441}, 429, 1494

\bibitem[\protect\citeauthoryear{Schneider et~al.,}{Schneider
  et~al.}{2010}]{schneider_sloan_2010}
Schneider D.~P.,  et~al., 2010, \mn@doi [The Astronomical Journal]
  {10.1088/0004-6256/139/6/2360}, 139, 2360

\bibitem[\protect\citeauthoryear{{Schweitzer} et~al.,}{{Schweitzer}
  et~al.}{2008}]{Schweitzer2008}
{Schweitzer} M.,  et~al., 2008, \mn@doi [\apj] {10.1086/587097}, \href
  {https://ui.adsabs.harvard.edu/abs/2008ApJ...679..101S} {679, 101}

\bibitem[\protect\citeauthoryear{{Shang} et~al.,}{{Shang}
  et~al.}{2011}]{shang_new_2011}
{Shang} Z.,  et~al., 2011, \mn@doi [\apjs] {10.1088/0067-0049/196/1/2}, \href
  {https://ui.adsabs.harvard.edu/abs/2011ApJS..196....2S} {196, 2}

\bibitem[\protect\citeauthoryear{Shen et~al.,}{Shen
  et~al.}{2011}]{shen_catalog_2011}
Shen Y.,  et~al., 2011, \mn@doi [The Astrophysical Journal Supplement Series]
  {10.1088/0067-0049/194/2/45}, 194, 45

\bibitem[\protect\citeauthoryear{{Shu}, {Wang}, {Yaqoob}, {Jiang}  \&
  {Zhou}}{{Shu} et~al.}{2012}]{Shu2012}
{Shu} X.~W.,  {Wang} J.~X.,  {Yaqoob} T.,  {Jiang} P.,   {Zhou} Y.~Y.,  2012,
  \mn@doi [\apjl] {10.1088/2041-8205/744/2/L21}, \href
  {https://ui.adsabs.harvard.edu/abs/2012ApJ...744L..21S} {744, L21}

\bibitem[\protect\citeauthoryear{{Stalevski}, {Fritz}, {Baes}, {Nakos}  \&
  {Popovic}}{{Stalevski} et~al.}{2012}]{stalevski_torus_2012}
{Stalevski} M.,  {Fritz} J.,  {Baes} M.,  {Nakos} T.,   {Popovic} L.~C.,  2012,
  Publications de l'Observatoire Astronomique de Beograd, \href
  {https://ui.adsabs.harvard.edu/abs/2012POBeo..91..235S} {91, 235}

\bibitem[\protect\citeauthoryear{Stalevski, Ricci, Ueda, Lira, Fritz  \&
  Baes}{Stalevski et~al.}{2016}]{stalevski_dust_2016}
Stalevski M.,  Ricci C.,  Ueda Y.,  Lira P.,  Fritz J.,   Baes M.,  2016,
  \mn@doi [Monthly Notices of the Royal Astronomical Society]
  {10.1093/mnras/stw444}, 458, 2288

\bibitem[\protect\citeauthoryear{{Stalevski}, {Tristram}  \&
  {Asmus}}{{Stalevski} et~al.}{2019}]{Stalevski2019}
{Stalevski} M.,  {Tristram} K. R.~W.,   {Asmus} D.,  2019, \mn@doi [\mnras]
  {10.1093/mnras/stz220}, \href
  {https://ui.adsabs.harvard.edu/abs/2019MNRAS.484.3334S} {484, 3334}

\bibitem[\protect\citeauthoryear{{Storchi-Bergmann} et~al.,}{{Storchi-Bergmann}
  et~al.}{2018}]{storchi_nlr_2018}
{Storchi-Bergmann} T.,  et~al., 2018, \mn@doi [\apj]
  {10.3847/1538-4357/aae7cd}, \href
  {https://ui.adsabs.harvard.edu/abs/2018ApJ...868...14S} {868, 14}

\bibitem[\protect\citeauthoryear{Toba et~al.,}{Toba
  et~al.}{2021}]{toba_how_2021}
Toba Y.,  et~al., 2021, \mn@doi [The Astrophysical Journal]
  {10.3847/1538-4357/abe94a}, 912, 91

\bibitem[\protect\citeauthoryear{{Ulrich}, {Maraschi}  \& {Urry}}{{Ulrich}
  et~al.}{1997}]{Ulrich1997}
{Ulrich} M.-H.,  {Maraschi} L.,   {Urry} C.~M.,  1997, \mn@doi [\araa]
  {10.1146/annurev.astro.35.1.445}, \href
  {https://ui.adsabs.harvard.edu/abs/1997ARA&A..35..445U} {35, 445}

\bibitem[\protect\citeauthoryear{{Urry} \& {Padovani}}{{Urry} \&
  {Padovani}}{1995}]{Urry1995}
{Urry} C.~M.,  {Padovani} P.,  1995, \mn@doi [\pasp] {10.1086/133630}, \href
  {https://ui.adsabs.harvard.edu/abs/1995PASP..107..803U} {107, 803}

\bibitem[\protect\citeauthoryear{{Wang}, {Xing}, {Zhang}, {Wang}, {Zhou}  \&
  {Zhang}}{{Wang} et~al.}{2013}]{wang_outflow_2013}
{Wang} H.,  {Xing} F.,  {Zhang} K.,  {Wang} T.,  {Zhou} H.,   {Zhang} S.,
  2013, \mn@doi [\apjl] {10.1088/2041-8205/776/1/L15}, \href
  {https://ui.adsabs.harvard.edu/abs/2013ApJ...776L..15W} {776, L15}

\bibitem[\protect\citeauthoryear{{Weedman}, {Sargsyan}, {Lebouteiller}, {Houck}
   \& {Barry}}{{Weedman} et~al.}{2012}]{Weedman2012}
{Weedman} D.,  {Sargsyan} L.,  {Lebouteiller} V.,  {Houck} J.,   {Barry} D.,
  2012, \mn@doi [\apj] {10.1088/0004-637X/761/2/184}, \href
  {https://ui.adsabs.harvard.edu/abs/2012ApJ...761..184W} {761, 184}

\bibitem[\protect\citeauthoryear{Whittaker}{Whittaker}{2009}]{whittaker_textbook_2009}
Whittaker J.,  2009, Graphical Models in Applied Multivariate Statistics.
Wiley Publishing

\bibitem[\protect\citeauthoryear{Wright et~al.,}{Wright
  et~al.}{2010}]{wright_wide-field_2010}
Wright E.~L.,  et~al., 2010, \mn@doi [The Astronomical Journal]
  {10.1088/0004-6256/140/6/1868}, 140, 1868

\bibitem[\protect\citeauthoryear{{Yang} et~al.,}{{Yang}
  et~al.}{2020}]{Yang2020}
{Yang} G.,  et~al., 2020, \mn@doi [\mnras] {10.1093/mnras/stz3001}, \href
  {https://ui.adsabs.harvard.edu/abs/2020MNRAS.491..740Y} {491, 740}

\bibitem[\protect\citeauthoryear{{Zhang}, {Dong}, {Wang}  \& {Gaskell}}{{Zhang}
  et~al.}{2011}]{zhang_baldwin_oiii_2011}
{Zhang} K.,  {Dong} X.-B.,  {Wang} T.-G.,   {Gaskell} C.~M.,  2011, \mn@doi
  [\apj] {10.1088/0004-637X/737/2/71}, \href
  {https://ui.adsabs.harvard.edu/abs/2011ApJ...737...71Z} {737, 71}

\bibitem[\protect\citeauthoryear{{Zhang}, {Wang}, {Gaskell}  \& {Dong}}{{Zhang}
  et~al.}{2013}]{Zhang2013}
{Zhang} K.,  {Wang} T.-G.,  {Gaskell} C.~M.,   {Dong} X.-B.,  2013, \mn@doi
  [\apj] {10.1088/0004-637X/762/1/51}, \href
  {https://ui.adsabs.harvard.edu/abs/2013ApJ...762...51Z} {762, 51}

\makeatother
\end{thebibliography}

\appendix

\section{The reliability of the r($\lambda$)}

The derived correlation coefficients between CF($\lambda$) and $\mathrm{EW}_{\mathrm{[OIII]5007}}$ show sharp rise toward shorter wavelength at $<$ 2 $\micron$ (see Fig. \ref{fig:curve}). This is simply
because that the significant contribution of the disc component to near infrared  could smear out the negative correlation.
Below we present simulations to verify. 

We first construct a mock sample of artificial CF0 and $\mathrm{EW}_{\mathrm{[OIII]5007}}$ with their statistical properties consistent with those of the real quasar sample but with a wavelength-independent intrinsic correlation coefficient between CF($\lambda$) and $\mathrm{EW}_{\mathrm{[OIII]5007}}$.
To do so, we first calculate mean vector and covariance matrix of
the four variables: CF0, EW, $L_{\mathrm{bol}}$ and $M_{\mathrm{BH}}$.

$$
\mu
=
\begin{bmatrix}
 \mu_1 \\
 \mu_2
\end{bmatrix}
\text{ with sizes }\begin{bmatrix} 2 \times 1 \\ 2 \times 1 \end{bmatrix}
$$

$$
\Sigma
=
\begin{bmatrix}
 \Sigma_{11} & \Sigma_{12} \\
 \Sigma_{21} & \Sigma_{22}
\end{bmatrix}
\text{ with sizes }\begin{bmatrix} 2 \times 2 & 2 \times 2 \\ 2 \times 2 & 2 \times 2 \end{bmatrix}
$$

where subscript 1 corresponds to CF0 and EW and subscript 2 corresponds to $L_{\mathrm{bol}}$ and $M_{\mathrm{BH}}$.

Inversion of a covariance matrix yields partial correlation matrix and vice versa \citep{whittaker_textbook_2009}. 
So we can modify the matrix element corresponding to CF0 and EW in the yielded partial correlation matrix and obtain covariance matrix with appointed partial correlation coefficient between CF0 and EW and keep correlations between other variables unchanged.
Using this new covariance matrix $\Sigma$, we construct a 4-dimensional 
log-normal distribution $N(\mu, \Sigma)$.
For a given source in the sample, $L_{\mathrm{bol}}$ and $M_{\mathrm{BH}}$ are given by observations (denoted as $\mathbf{a}$), its simulated CF0 and EW are sampled from the conditional distribution $N(\bar{\mu},\bar{\Sigma})$, where
$
\bar{\mu} = \mu_1 + \Sigma_{12} \Sigma_{22}^{-1} \left( \mathbf{a} - \mu_2 \right),
\bar{\Sigma} = \Sigma_{11}-\Sigma_{12}\Sigma_{22}^{-1}\Sigma_{21}
$ \citep{eaton_multivariate_1983}.
The simulated CF0 and EW are thus guaranteed to have distributions (mean and scatter) consistent with those of the real sample (CF0 from the hot+cold dust model fitting, and EW from SDSS spectra), but with assigned intrinsic correlation between them.

We could then build mock SED data points for each source with the simulated CF0 and EW. We adjust the normalization of the mean dust component (see Fig. \ref{fig:polar}) to match the simulated CF0 for each source, and add the dust component to the best-fit disk component derived from the hot+cold dust model fitting.
Convolving the mock SED model with the transmission curves of photometric bands, and adding the residuals obtained through fitting the real SED (see Fig. \ref{fig:example}), we build mock SED data points for each source. 

\begin{figure}
    \centering
    \includegraphics[width=0.45\textwidth]{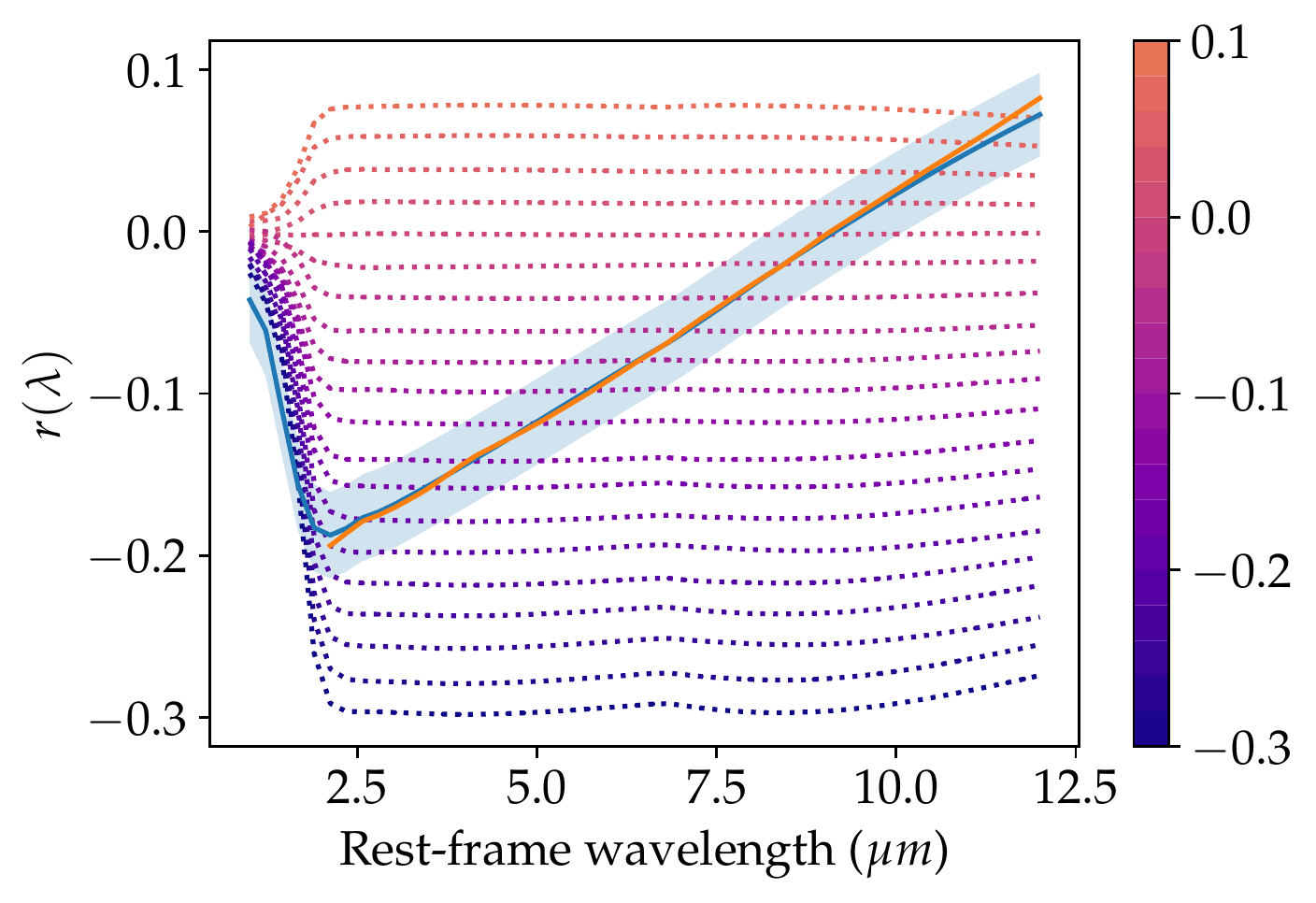}
    \caption{Monte-Carlo simulations show how the mock wavelength-independent correlation coefficients (with 20 input values varying from -0.3 to 0.08, with a step size of 0.02) could be recovered (color coded dotted lines) in mock samples. The blue line plots the $r$($\lambda$) derived from the real sample, while the the yellow line shows the corrected $r$($\lambda$) after registering with the simulated $r$($\lambda$). Clearly, such correction is negligible (compared with the statistical uncertainties which are plotted as blue shadows ) at $\lambda$ $>$ 2.1$\micron$, while at shorter wavelengths reliable correction would be infeasible. 
    }
    \label{fig:corrected}
\end{figure}

Repeating the SED interpolation and correlation analyses, we could derive $r$($\lambda$) for the mock sample. %
For various input correlation coefficients, the output $r$($\lambda$) derived from the mock samples quickly converge to zero at $\lambda$ $<$ 2 $\micron$.
Comparing the $r$($\lambda$) of the real and mock sample, we can claim
that the rise of the observed $r$($\lambda$) with decreasing wavelength at $\lambda$ smaller than $\sim$ 2.1 $\micron$ is clearly an artifact. This is because at such short wavelengths, the contribution to total photometry from the disc component could outshine the dust component (see Fig. \ref{fig:example}), which could subsequently 
smear out the anti-correlation between CF($\lambda$) and EW. 
The $r$($\lambda$) from the mock samples appear flat (though subtle curvatures are visible) at $\lambda$ $>$ 2.1 $\micron$, and the input wavelength-independent correlation coefficients $r$ can be well recovered.
The subtle curvatures at $\lambda$ $>$ 2.1 $\micron$ are due to weak biases in our SED analyses (interpolation could yield extra scatter) and the wavelength-dependent photometric uncertainties (see Fig. \ref{fig:residual}). However such biases are negligible in this work.

\bsp	%
\label{lastpage}
\end{document}